\documentclass[journal]{IEEEtran}
\usepackage{amsmath,amsfonts}
\usepackage{array}
\usepackage{textcomp}
\usepackage{stfloats}
\usepackage{url}
\usepackage{verbatim}
\usepackage{graphicx}
\usepackage{cite}
\usepackage{CJK}
\usepackage{subfigure}
\usepackage[linesnumbered, ruled]{algorithm2e}
\SetKwRepeat{Do}{do}{while}

\begin{document}

\title{Over-the-Air Diagnosis of Defective Elements in Intelligent Reflecting Surface}
 \author{%
	Ziyi~Zhao,
	Zhaorui~Wang,
	Lin~Zhou,
	Chunsong~Sun,
	Shuowen~Zhang,
	Naofal~Al-Dhahir,
	and Liang Liu
	\thanks{This paper was presented in part at the IEEE Global Conference of Communications (GLOBECOM) 2024 \cite{Zhao24}.}
	\thanks{Z. Zhao, S. Zhang, and L. Liu are with the Department of Electrical and Electronic Engineering, The Hong Kong Polytechnic University, Hong Kong SAR, China (e-mails: ziyii.zhao@connect.polyu.hk, \{shuowen.zhang, liang-eie.liu\}@polyu.edu.hk).}
	\thanks{Z. Wang is with the FNii and SSE, The Chinese University of Hong Kong, Shenzhen, China (e-mail: wangzhaorui@cuhk.edu.cn).}
	\thanks{L. Zhou and C. Sun are with the School of Cyber Science and Technology, Beihang University, Beijing, China, 100191 (e-mails: \{lzhou, sunchunsong\}@buaa.edu.cn).}
	\thanks{N. Al-Dhahir is with the Department of Electrical and Computer Engineering, University of Texas at Dallas, Richardson, USA (email: aldhahir@utdallas.edu).}
}

\maketitle

\newtheorem{remark}{Remark}
\newtheorem{example}{Example}
\begin{abstract}
Due to circuit failures, defective elements that cannot adaptively adjust the phase shifts of their impinging signals in a desired manner may exist on an intelligent reflecting surface (IRS). Traditional way to locate these defective IRS elements requires a thorough diagnosis of all the circuits belonging to a huge number of IRS elements, which is practically challenging. In this paper, we will devise novel approaches under which a transmitter sends known pilot signals and a receiver localizes all the defective IRS elements just based on its over-the-air measurements reflected from the IRS. Specifically, given any set of IRS elements, we propose an efficient method to process the received signals to determine whether this cluster contains defective elements or not with a very high accuracy probability. Based on this method, we show that the over-the-air diagnosis problem belongs to the 20 questions problem, where we can adaptively change the query set at the IRS so as to localize all the defective elements as quickly as possible. Along this line, we first propose a sorted posterior matching (sortPM) based method according to the noisy 20 questions technique, which enables accurate diagnosis even if the answers about the existence of defective elements in some sets of interest are wrong at certain question and answer (Q\&A) rounds due to the noisy received signals. Next, to reduce the complexity, we propose a bisection based method according to the noiseless 20 questions technique, which totally trusts the answer at each Q\&A round and keeps removing half of the remaining region based on such answers. Via numerical results, we show that our proposed methods can exploit the over-the-air measurements to localize all the defective IRS elements quickly and accurately.
\end{abstract}


\section{Introduction}
\subsection{Motivation}
\IEEEPARstart{T}{H}E recent revolution in software-controlled surfaces using metamaterials has stimulated a flurry of research activities in using intelligent reflecting surface (IRS) to improve the performance for wireless communication \cite{Liaskos18,Renzo19,Basar19,Qingqing21}. The key is to adaptively adjust the phase shifts of the IRS elements based on the channel state information (CSI) so as to enhance the channel quality of the mobile users. Previously, tremendous works have been done to enable low-overhead CSI acquisition \cite{wang2020channel, liu2020matrix,Zhou22,Chen23,Runnan24,Rui25}. Moreover, various works have designed efficient IRS beamforming algorithms to utilize the CSI for optimizing the communication performance in IRS-assisted systems \cite{Wu19IRS}\cite{Mu20IRS}. Due to the importance of integrated sensing and communication (ISAC) in 6G \cite{Liu24ISAC,Zheng19ISAC,Liuan22ISAC}, the IRS beamforming design has also been optimized to balance between the sensing and communication performance \cite{Song2022ISAC,Shao2022ISAC,Fang2024ISAC}. \par
Note that in practice, the circuits of IRS elements are prone to failures, because a number of electrical circuits and tunable elements are densely integrated in the programmable metasurface \cite{Srinivasan04,Saeed18,Taghvaee20}. For example, when the connection between a controller chip and its associated IRS elements is broken, this controller cannot send the desired phase shifts to these defective elements dynamically, whose phase shifts will be stuck at some fixed states \cite{Saeed18}. Furthermore, each row or each column of the IRS elements is usually connected to one common power supply circuit. Therefore, if one power supply circuit does not work, a whole row or column of the IRS elements will not function \cite{Taghvaee20}. To summarize, after the IRS beamforming vector is designed based on the algorithms in prior works, it is very likely that some defective elements exist on the IRS and they cannot adaptively achieve the desired reflecting patterns. This calls for an efficient method for finding the defective IRS elements to fully reap the IRS beamforming design gain.\par
\begin{figure}
	\centering
	\includegraphics[scale=0.8]{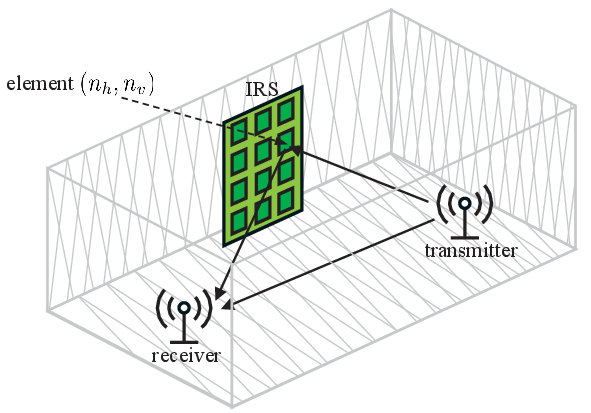}
	\caption{The over-the-air IRS diagnosis system in an anechoic chamber.}
	\label{fig.1}
\end{figure}

A conventional approach for a diagnosis of the IRS defective elements is to have a thorough check of all the circuits. However, this approach is practically challenging, because there are a huge number of sophisticated circuits to control hundreds of or even thousands of IRS elements. To avoid exhaustive circuit check with prohibitive complexity, this paper is interested in the over-the-air diagnosis method. Specifically, as shown in Fig. \ref{fig.1}, a radio transmitter, a radio receiver, and the IRS to be checked are deployed in an anechoic chamber. Because the signals at the receiver side over the transmitter-IRS-receiver channel will be different from the expected ones when some defective IRS elements do not reflect the signals in the desired manner, we hope that via carefully adjusting the IRS reflecting coefficients and observing the corresponding over-the-air measurements at the receiver, we can find all the defective IRS elements based on some signal processing technique.
\subsection{Prior Work} 
It is worth noting that a related topic, IRS hardware impairment, has been investigated in the literature \cite{Badiu20, Khel22, Alouini21, Zhang22}. Specifically, due to the imperfect hardware, it is assumed that an unknown but fixed phase error is added to the desired phase shift of each IRS element. Based on this model, various signal processing techniques have been proposed to estimate these phase errors such that calibration can be conducted to compensate for these phase errors. However, circuit failure considered in this paper is quite different from hardware impairment considered in the above works. Specifically, under the hardware impairment model, the true phase shifts of all IRS elements are linear functions of the desired phase shifts, with additive phase errors. However, under the circuit failure model considered in this paper, some IRS elements are totally in the irregular state, and the phase shifts of these defective elements usually have nothing to do with the desired values, e.g., they are stuck at fixed values. There are two differences over calibration. First, the locations of the defective elements are unknown and need to be estimated. Second, the true phase shifts of the defective elements are no longer linear functions of the desired phase shifts, and it is not possible to utilize the linear techniques that are widely used in calibration. Due to the above reasons, the calibration technique does not work under our considered model. We have to localize all the defective IRS elements, even though we do not know their irregular reflecting patterns, such that we can repair their circuits. This is a new problem.\par
Another related topic in the literature is the over-the-air antenna array diagnosis \cite{Bucci05,Migliore11,Oliveri12,Fuchs16,Xiong19}. In these works, it is assumed that some fault antennas exist at the transmitter side and they cannot emit radio signals. Mathematically, the beamforming coefficients of the defective antennas are stuck at zero. Because the desired beamforming coefficient and the real beamforming coefficient (i.e., 0) of each defective antenna is known, the difference between the desired and true received signals is a known response of the defective antennas. By assuming that the number of defective antennas is small, the compressed sensing technique is used to localize the defective antennas. However, the above techniques do not work in our considered IRS diagnosis problem. First, in the above works, the beamforming coefficients of defective antennas are stuck at a known state zero. However, this is not true in our problem. For example, when the connection between a controller and its associated IRS elements is broken, all these defective elements will be stuck at the states set at the last moment when the connections works. Because we do not know when the connection is broken, we have no knowledge about the stuck states of the defective IRS elements. In this case, the difference between the desired and true received signals is no longer a known response of the defective IRS elements. Second, in this work, we do not limit the number of defective elements to be small. 
\subsection{Main Contributions}
In this paper, we target at localizing all the defective elements in an IRS. The contributions of this paper are summarized as follows.
\begin{itemize}
	\item First, in sharp contrary to the conventional approach under which the complex circuits of all IRS elements have to be thoroughly checked, we propose a novel over-the-air method for diagnosis of the IRS. Specifically, a transmitter emits known pilot signals while a receiver processes its over-the-air measurements reflected by the IRS to localize the defective IRS elements, as shown in Fig. \ref{fig.1}. This approach is feasible because the received signals are functions of the set of the IRS defective elements, and via a proper design of the desired reflecting coefficients of all the IRS elements over time, we are able to perform IRS diagnosis based on the abnormal state in the received signals. Such an over-the-air approach can significantly simplify IRS diagnosis in practice.
\end{itemize}	
\begin{itemize}
	\item Second, we show that the IRS defective elements diagnosis problem is actually a 20 questions problem \cite{Renyi1984,Ulam1991,Schalk1971,Hill1992,Pelc02}. In a 20 questions game, a responder has to answer the questions raised by the questioner (sometimes, the responder may also lie to the questioner). Based on the answers, the job of the questioner is to adaptively design the questions such that the right solution can be obtained via as fewer queries as possible. Under our considered IRS diagnosis problem, an IRS reflecting coefficient control strategy is devised such that given any query set of IRS elements, we manage to know whether any defective elements exist in this set or not with a high accuracy probability, just based on the over-the-air observations at the receiver. Because of this capacity, we point out that the challenge to utilize the 20 questions technique lies in how to design the query set of IRS elements adaptively so as to localize all defective elements quickly.
\end{itemize}
\begin{itemize}
	\item Third, we utilize a noisy 20 questions technique, the sortPM algorithm \cite{Chiu16,Chiu21,Chiu19}, to address the above problem. Under the sortPM algorithm, we aim to determine the boundary of the defective IRS element set. Specifically, based on the over-the-air measurements at each time slot, we first update the posterior probabilities of all the values to be the boundary point, then sort these probabilities in an descending order, and last set the query set in the next question and answer (Q\&A) round based on the values with the largest probabilities. Because the probabilities of all the points to become the boundary are recorded at each time slot, the sortPM algorithm has the ability to localize the defective elements even when some answers about the existence of defective elements in the query set are wrong. 
\end{itemize}
\begin{itemize}
	\item Fourth, to further reduce the computational complexity, we also utilize a noiseless 20 questions technique, the bisection method, for IRS defective element diagnosis. Under this approach, we assume that the response based on the over-the-air measurements is always correct, and remove half of the remaining region that does not contain the boundary point at each time slot. Numerical results show that when the signal-to-noise ratio (SNR) is high or when the number of antennas at the receiver is large, the bisection method can achieve similar performance to the sortPM method, but with a much lower complexity.
\end{itemize}		
\subsection{Organization}
The rest of this paper is organized as follows. Section II describes the system model for over-the-air diagnosis of defective IRS elements. Section III and Section IV introduce the noisy and noiseless 20 questions techniques to localize the defective IRS elements, respectively. Section V provides the numerical results pertaining to the diagnosis performance. Finally, Section VI concludes this paper.\par
\section{System Model}
Consider an IRS consisting of $N=N_h\times N_v$ elements, which are laid out in a grid pattern with $N_h$ columns and $N_v$ rows. Without loss of generality, we assume that $N_h=2^{m_h}$ and $N_v=2^{m_v}$ with some integers $m_h$ and $m_v$. Define $\mathcal{N}_h=\{1,\dots,N_h\}$ and $\mathcal{N}_v=\{1,\dots,N_v\}$. Moreover, we index the element at the $n_h$-th column and the $n_v$-th row of the IRS as element $(n_h,n_v)$, $n_h\in\mathcal{N}_h$, $n_v\in\mathcal{N}_v$.\par
Because of the complicated circuit to control the reflecting patterns, the elements on an IRS are prone to failures. The defective IRS elements may not reflect the signals in the pre-designed pattern, leading to poor communication performance. If an IRS is in an irregular state with some defective elements, our objective is to find the indices of all the defective elements.\par
\subsection{IRS Failure Model}
\begin{figure}
	\centering
	\includegraphics[scale=0.55]{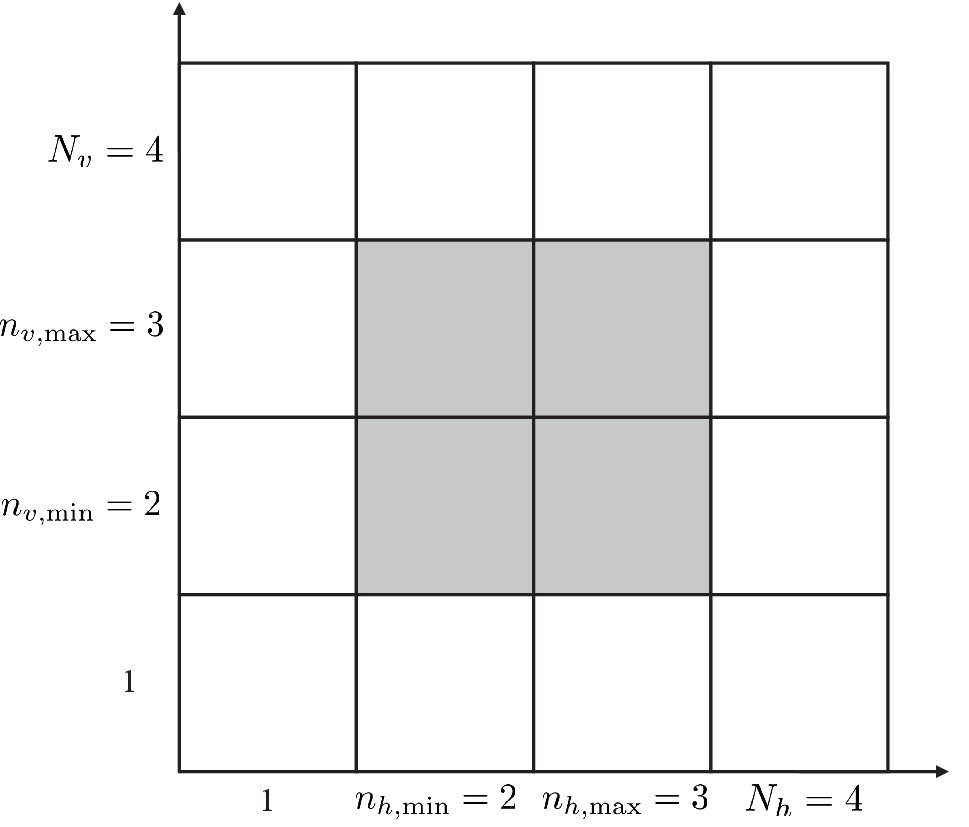}
	\caption{An IRS with clustered defective IRS elements (in gray): there are $N_v=4$ rows and $N_h=4$ columns of elements on the IRS, and the indices of the defective elements are $(2,2)$, $(2,3)$, $(3,2)$ and $(3,3)$.}
	\label{fig.2}
\end{figure}
In this paper, we consider a clustered failure model of the IRS, where all the defective elements are located within a continuous rectangular region of the IRS,\footnote{If the defective elements are located within a continuous but non-rectangular region of the IRS, the defective element set $\mathcal{E}$ defined in (\ref{eqn:defective element set}) actually is the minimum rectangular region that contains all the defective elements. After such a small region is found based on the over-the-air approach that is later proposed in the paper, we can have a quick diagnosis of circuits of all the elements in $\mathcal{E}$ to further localize the defective elements.} as shown in Fig. \ref{fig.2}. Such a model is valid in many scenarios, e.g., when a controller chip fails to send control signals to its associated elements, when faults in some interconnects leave a clustered region of the IRS isolated from the remaining region, etc. Let $(n_{h,{\rm min}}, n_{v,{\rm min}})$ and $(n_{h,{\rm max}},n_{v,{\rm max}})$ denote the indices of the elements at the lower left corner and the upper right corner of this defective region, respectively, as shown in Fig. \ref{fig.1}. Therefore, under our considered clustered failure model, the set consisting of the indices of all the defective elements on the IRS is given by
\begin{align}\label{eqn:defective element set}
    \mathcal{E}=\{(n_h,n_v)|&n_h\in\{n_{h,{\rm min}},\dots,n_{h,{\rm max}}\}, \nonumber \\
    &n_v\in\{n_{v,{\rm min}},\dots,n_{v,{\rm max}}\}\}.
\end{align}
For simplicity, we also define the set consisting of the indices of all the IRS elements in the normal state as
\begin{align}\label{eqn:normal element set}
    \mathcal{W}=\{(n_h,n_v)|(n_h,n_v)\notin\mathcal{E}\}.
\end{align}

Let $\phi_{n_h,n_v,t}\in (0,2\pi]$ and $e^{j\phi_{n_h,n_v,t}}$ denote the desired phase shift and the corresponding reflecting coefficient of IRS element $(n_h,n_v)$ at time slot $t$, $\forall n_h\in\mathcal{N}_h$, $n_v\in\mathcal{N}_v$, $t=1,...,T$, where $T$ denotes the number of time slots that we consider. For each IRS element $(n_h,n_v)\in\mathcal{W}$ that is in the normal state, it will adaptively reflect the signals with different phase shifts at different time slots. However, for any defective IRS element $(n_h,n_v)\in\mathcal{E}$, we assume that it is stuck at a constant but unknown state over different time slots. This can occur when a controller chip fails to send control signals to its associated elements, such that these elements cannot adaptively update their reflecting coefficients as desired \cite{Saeed18}. Under this error pattern, the constant phase shift and the corresponding reflecting coefficient of a defective element $(n_h,n_v)\in\mathcal{E}$ across all the time slots are defined as $\beta_{n_h,n_v}\in(0,2\pi]$ and $e^{j\beta_{n_h,n_v}}$, respectively. In practice, the real reflecting coefficient of IRS element $(n_h,n_v)$ at time slot $t$, which depends on whether this element is defective or not, is given by
\begin{align}\label{eqn:reflecting coefficient}
\theta_{n_h,n_v,t}=\left\{\begin{array}{ll} e^{j\phi_{n_h,n_v,t}}, & {\rm if} ~ (n_h,n_v)\in \mathcal{W}, \\ e^{j\beta_{n_h,n_v}}, & {\rm if} ~ (n_h,n_v)\in \mathcal{E},\end{array}\right. ~ \forall t.
\end{align}
Note that under the above reflecting model, $\beta_{n_h,n_v}$ and $\mathcal{E}$ (also $\mathcal{W}$) are unknown. Therefore, $\theta_{n_h,n_v,t}$ is also unknown, $\forall n_h, n_v, t$.
\subsection{Signal Model}
In this paper, we aim to propose an over-the-air approach to estimate the values of $n_{h,{\rm min}}$, $n_{h,{\rm max}}$, $n_{v,{\rm min}}$, and $n_{v,{\rm max}}$ such that we can find the indices of all the defective elements on the IRS, i.e., $\mathcal{E}$. To enable the over-the-air diagnosis of the IRS, we deploy a radio transmitter, a radio receiver, and the IRS in an anechoic chamber, as shown in Fig. \ref{fig.1}. It is assumed that the transmitter is equipped with 1 antenna, while the receiver is equipped with $M\geq 1$ antennas. Define $\boldsymbol{h}\in\mathbb{C}^{M\times 1}$ as the channel vector from the transmitter to the receiver, $u_{n_h,n_v}$ as the channel coefficient from the transmitter to IRS element $(n_h,n_v)$, and $\boldsymbol{r}_{n_h,n_v}\in\mathbb{C}^{M\times 1}$ as the channel vector from IRS element $(n_h,n_v)$ to the receiver, $\forall n_h, n_v$. Moreover, define $\boldsymbol{g}_{n_h,n_v}=u_{n_h,n_v}\boldsymbol{r}_{n_h,n_v}$ as the cascaded channel from the transmitter to IRS element $(n_h,n_v)$ to the receiver. At time slot $t$, the received signal at the receiver side is then given by
\begin{align}\label{eqn:received signal general}
\boldsymbol{y}_t=&\boldsymbol{h}x_t+\sum_{(n_h,n_v)\in\mathcal{E}}e^{j\beta_{n_h,n_v}}\boldsymbol{g}_{n_h,n_v}x_t \nonumber \\ &+\sum_{(n_h,n_v)\in\mathcal{W}}e^{j\phi_{n_h,n_v,t}}\boldsymbol{g}_{n_h,n_v}x_t+\boldsymbol{z}_t,~ t=1,\dots,T,
\end{align}
where $x_t$ denotes the known pilot transmitted at time slot $t$, and $\boldsymbol{z}_t\sim\mathcal{CN}(\boldsymbol{0},\sigma^2\boldsymbol{I})$ denotes the additive white Gaussian noise (AWGN) of the receiver at time slot $t$. \par
\subsection{Brief Introduction of Overall Over-the-Air Approach}
Note that the received signals given in (\ref{eqn:received signal general}) are functions of the unknown set $\mathcal{E}$. Under the over-the-air approach, we will dynamically adjust the desirable phase shifts of all the IRS elements, i.e., $\phi_{n_h,n_v,t}$'s, and the transmit signals, i.e., $x_t$'s, over time so as to receive time-varying received signals of $\boldsymbol{y}_1,\dots,\boldsymbol{y}_T$. Then, based on $T$ different equations related to the over-the-air received signals as shown in (\ref{eqn:received signal general}), we aim to estimate the locations of the defective IRS elements. \par
A brief description for our approach to estimate $\mathcal{E}$ based on $\boldsymbol{y}_t$'s is as follows. In this paper, we estimate $n_{h,{\rm min}}$, $n_{v,{\rm min}}$, $n_{h,{\rm max}}$, and $n_{v,{\rm max}}$, individually, because the defective element set is defined by these four points, as shown in (\ref{eqn:defective element set}). As will be shown in Section III, in this paper, we design a novel approach to control $\phi_{n_h,n_v,t}$'s and $x_t$'s such that given any query set, we know whether $n_{h,{\rm min}}/n_{v,{\rm min}}/n_{h,{\rm max}}/n_{v,{\rm max}}$ is in this set correctly with a very high probability. Then, we keep updating the query set until we are sure that one element in this set is $n_{h,{\rm min}}/n_{v,{\rm min}}/n_{h,{\rm max}}/n_{v,{\rm max}}$ with sufficiently high probability. Therefore, our problem reduces to the 20 questions problem \cite{Ulam1991}, which aims to guess an integer via asking a minimum number of questions. \par
In the rest of this paper, we design two methods for localizing the defective IRS elements. First, because given any query set, we cannot know whether $n_{h,{\rm min}}/n_{v,{\rm min}}/n_{h,{\rm max}}/n_{v,{\rm max}}$ is in this set with probability 1 due to the noise in the received signals, Section III will present a noisy 20 questions problem based method \cite{Pelc02}. Second, to reduce the complexity of the noisy 20 questions technique, in Section IV, we will introduce a noiseless 20 questions problem based algorithm \cite{Renyi1984}, i.e., the bisection algorithm, by assuming that our decision about whether $n_{h,{\rm min}}/n_{v,{\rm min}}/n_{h,{\rm max}}/n_{v,{\rm max}}$ is in the query set is always correct. \par

\section{Sorted Posterior Matching based Method}
In this section, we introduce a sortPM-based method\cite{Chiu16,Chiu21,Chiu19}, a noisy 20 questions technique, for detecting the defective elements on the IRS. In the following, we first briefly introduce the sortPM algorithm, and then show how to apply it to our considered IRS diagnosis problem. 
\subsection{Introduction of SortPM Algorithm}
The sortPM method is a powerful technique to solve the noisy 20 questions problem, which is described as follows. Suppose a responder thinks of an integer, denoted by $\omega$ from the set $\{1,2,\dots,D\}$, where $D$ is an integer. In the $k$-th round of Q\&A, the questioner selects a query set $\mathcal{S}_k\subseteq\{1,2,\dots,D\}$, inspects for the presence of $\omega$ in this set, and then gets a binary yes-or-no answer from the responder, where $Y_k=1$ indicates yes and $Y_k=0$ indicates no. Specifically, the responder is allowed to lie with a fixed probability $q\in(0, 0.5)$, i.e., the probability of telling the truth is larger than that of lying. Define the true answer indicator variable at the $k$-th Q\&A round as
\begin{align}\label{eqn:X_k}
	X_{k}=\left\{\begin{array}{ll} 
		1, & {\rm if} ~\omega\in\mathcal{S}_k, \\ 
		0, & {\rm if} ~\omega\notin\mathcal{S}_k, \end{array}\right.
		\forall k. 
\end{align}
Then, the noisy binary answers can be modeled as
\begin{align}\label{eqn:Y_k} 
	Y_k = X_k\oplus Z_k, \forall k,
\end{align}
where $\oplus$ denotes the exclusive OR operation, and $Z_k\sim{\rm Bern}(q)$ denotes the binary noise in the $k$-th round. \par
Under this lying pattern, the objective of the questioner is to adaptively design its query sets based on the answers from the responder so as to correctly guess $\omega$ through a minimum number of questions. In the following, we introduce the key component of the sortPM method - given the answers from the responder in the above $k-1$ Q\&A rounds, i.e., $\boldsymbol{Y}^{k-1}=\{Y_1,\dots,Y_{k-1}\}$, how to design the query set asked in the $k$-th Q\&A round, i.e., $\mathcal{S}_k, \forall k$. \par
Define
\begin{align}\label{eqn:pi i k}
     \pi_d(k-1)=P({\omega}=d|\boldsymbol{Y}^{k-1}),~d=1,2,\dots,D,~\forall k,
\end{align}
as the posterior probability that the true integer is $d$ given the answers $\boldsymbol{Y}^{k-1}$ obtained at the previous $k-1$ rounds of Q\&A. Based on the Bayesian technique, after the $(k-1)$-th Q\&A round, $\pi_d(k-1),~d=1,2,\dots,D,$ can be obtained recursively as follows 
\begin{align}\label{eqn:pi i k Bayes}
	\pi_d(k-1)&=\frac{P(\omega=d,\boldsymbol{Y}^{k-1})}{P(\boldsymbol{Y}^{k-1})} \nonumber \\
	&=\frac{P(\omega=d,Y_{k-1},\boldsymbol{Y}^{k-2})}{P(Y_{k-1},\boldsymbol{Y}^{k-2})} \nonumber \\
	&=\frac{P(Y_{k-1}|\omega=d,\boldsymbol{Y}^{k-2})P(\omega=d|\boldsymbol{Y}^{k-2})}{P(Y_{k-1}|\boldsymbol{Y}^{k-2})} \nonumber \\
	&=\frac{P(Y_{k-1}|\omega=d,\boldsymbol{Y}^{k-2})P(\omega=d|\boldsymbol{Y}^{k-2})}{\sum_{j=1}^{D}P(Y_{k-1}|\omega=j, \boldsymbol{Y}^{k-2})P(\omega=j|\boldsymbol{Y}^{k-2})}.
\end{align}
Note that given $\forall j\in\mathcal{S}_{k-1}$, it follows that $P(Y_{k-1}=1|\omega=j, \boldsymbol{Y}^{k-2})=1-q$, i.e., the responder gives the correct answer that $\omega$ is in the set $\mathcal{S}_{k-1}$, and $P(Y_{k-1}=0|\omega=j, \boldsymbol{Y}^{k-2})=q$, i.e., the responder lies that $\omega$ is not in the set $\mathcal{S}_{k-1}$. Similarly, given $\forall j\notin\mathcal{S}_{k-1}$, it follows that $P(Y_{k-1}=0|\omega=j, \boldsymbol{Y}^{k-2})=1-q$, and $P(Y_{k-1}=1|\omega=j, \boldsymbol{Y}^{k-2})=q$. Moreover, $P(\omega=j|\boldsymbol{Y}^{k-2})=\pi_d(k-2)$. Therefore, $\pi_d(k-1)$ given in (\ref{eqn:pi i k Bayes}) reduces to (\ref{eqn:pi k-1 1}) and (\ref{eqn:pi k-1 0}) on the top of this page. 
\begin{figure*}
	\begin{align}\label{eqn:pi k-1 1}
		\pi_{d}(k-1)=\left\{\begin{array}{ll}
			\frac{(1-q)*\pi_{d}(k-2)}{\sum_{j\in\mathcal{S}_{k-1}}(1-q)*\pi_{j}(k-2)+\sum_{j\notin\mathcal{S}_{k-1}}q*\pi_{j}(k-2)}, &\forall d\in\mathcal{S}_{k-1}, \\
			\frac{q*\pi_{d}(k-2)}{\sum_{j\in\mathcal{S}_{k-1}}(1-q)*\pi_{j}(k-2)+\sum_{j\notin\mathcal{S}_{k-1}}q*\pi_{j}(k-2)}, &\forall d\notin\mathcal{S}_{k-1},    
		\end{array} \right. 
		{\rm if}~Y_{k-1}=1.
	\end{align}
	\hrule
\end{figure*}
\begin{figure*}
	\begin{align}\label{eqn:pi k-1 0}
		\pi_{d}(k-1)=\left\{\begin{array}{ll}
			\frac{q*\pi_{d}(k-2)}{\sum_{j\in\mathcal{S}_{k-1}}q*\pi_{j}(k-2)+\sum_{j\notin\mathcal{S}_{k-1}}(1-q)*\pi_{j}(k-2)}, &\forall d\in\mathcal{S}_{k-1}, \\
			\frac{(1-q)*\pi_{d}(k-2)}{\sum_{j\in\mathcal{S}_{k-1}}q*\pi_{j}(k-2)+\sum_{j\notin\mathcal{S}_{k-1}}(1-q)*\pi_{j}(k-2)}, &\forall d\notin\mathcal{S}_{k-1},    
		\end{array} \right. 
		{\rm if}~Y_{k-1}=0.
	\end{align}
	\hrule
\end{figure*}

Let $\boldsymbol{\pi}(k-1)=[\pi_1(k-1),\dots,\pi_{D}(k-1)]$ denote the posterior probability vector at the $(k-1)$-th Q\&A round, $\forall k$, based on (\ref{eqn:pi k-1 1}) and (\ref{eqn:pi k-1 0}). 
For simplicity, we also define the sorted posterior probability vector of $\omega$ in the $(k-1)$-th round as 
\begin{align}\label{eqn:pi k-1 sorted}
	\boldsymbol{\pi}^{\downarrow}(k-1)=[\pi^{\downarrow}_1(k-1),\dots,\pi^{\downarrow}_d(k-1),\dots,\pi^{\downarrow}_D(k-1)],
\end{align}
where $\pi^{\downarrow}_d(k-1)$ denotes the $d$-th largest element in $\boldsymbol{\pi}(k-1)$, $\forall d$, such that all the elements in $\boldsymbol{\pi}^{\downarrow}(k-1)$ are in a descending order. Define $\gamma_{1}(k-1),\dots,\gamma_{D}(k-1),$ as the sorting operation where $\pi^{\downarrow}_d(k-1) = \pi_{\gamma_{d}(k-1)}(k-1)$. Therefore, in $\{1,\dots,D\}$,
element $\gamma_{d}(k-1)$ denotes the integer such that after obtaining the answer $Y_{k-1}$ in the $(k-1)$-th Q\&A round, $P(\omega=\gamma_{d}(k-1)|Y_{k-1})$ is the $d$-th largest posterior probability in the set $\{\pi_d(k-1)|d=1,\dots,D\}$.\par
Under the sortPM method, at the $k$-th Q\&A round, we can first obtain the sorted posterior probability vector $\boldsymbol{\pi}^{\downarrow}(k-1)$ according to (\ref{eqn:pi k-1 sorted}), and set the query set as
\begin{align}\label{eqn:query set sortPM}
	\mathcal{S}_{k}=\{\gamma_{1}(k-1),\dots,\gamma_{l^\ast}(k-1)\},~\forall k,
\end{align}
where
\begin{align}\label{eqn:l star sortPM}
	l^\ast=\mathop{\arg\min}\limits_{l} \left| \sum_{d=1}^{l}\pi^{\downarrow}_{d}(k-1)-\frac{1}{2}\right|.
\end{align} 
Thereby, based on the answer ${Y}_{k-1}$, at the $k$-th Q\&A round, the questioner selects the first $l^{\ast}$ largest posterior probabilities in $\{\pi^{\downarrow}_d(k-1)|d=1,\dots,D\}$ to construct the query set. Here, $l^\ast$ is selected such that given ${Y}_{k-1}$, the probability of the event $\omega\in\mathcal{S}_k$ at the $k$-th Q\&A round is closest to 0.5. We use this criterion to select $l^\ast$ because the entropy function of the binary random variable $X_k$ attains the maximum value when $p(X_k=1)=1/2$\cite{Chiu21}.\par 
To summarize, we can keep updating the query set based on (\ref{eqn:query set sortPM}) and (\ref{eqn:l star sortPM}) to ask whether $\omega$ is in this set, until at some round $k$, the largest probability among all guesses, i.e., $\pi_{1}^{\downarrow}(k)$, is already very close to 1. Now, we can guess $\omega=\gamma_{1}(k)$, because $P(\omega=\gamma_{1}(k)|\boldsymbol{Y}^{k})$ is very close to 1. The sortPM algorithm is summarized in Algorithm 1. \par
Note that under the sortPM algorithm, at each Q\&A round $k$, we update the posterior probabilities of all the possible values of $\omega$, i.e., $\pi_1(k),\dots,\pi_D(k),$ according to (\ref{eqn:pi k-1 1}) and (\ref{eqn:pi k-1 0}). In other words, we treat all integers as the possible solution even if some posterior probabilities are very small at certain Q\&A round. Therefore, under the sortPM method, even if some lies can make the posterior probability of the true integer small at certain Q\&A rounds, we still keep updating its posterior probability. After more right answers are provided by the responder, eventually, the true integer will yield the maximum posterior probability and be selected. An example is provided as follows to verify the ability of the sortPM method to make the right guess with lies. 
\begin{algorithm}
   \caption{Sorted Posterior Matching Method}
   \label{alg:sortPM}
  {\bf Input}: integer $D$, probability of lie $q$, and a positive threshold $\epsilon$ close to $0$\;
  {\bf Initialization}: $\pi_{d}(0)=1/D$, $\forall d=1,2,\dots,D$\;
  \For{$k=1,2,\dots$}
  { 
  Design the query set $\mathcal{S}_{k}$ according to (\ref{eqn:query set sortPM}) and (\ref{eqn:l star sortPM})\; 
  Ask whether $\omega\in\mathcal{S}_{k}$ and get an answer $Y_{k}$\;
  Update ${\pi}_{d}(k),~\forall d=1,2,\dots,D,$ based on ${Y}_{k}$ according to (\ref{eqn:pi k-1 1}) and (\ref{eqn:pi k-1 0})\;
  \If{$\pi_{1}^{\downarrow}(k)\geq 1-\epsilon$}
  {break\;} 
  }
  The estimation of $\omega$: $\hat{\omega}=\gamma_{1}(k)$\;
  {\bf Output}: $\hat{\omega}$\;
\end{algorithm}
\begin{example}
Suppose we want to guess the number $1$ from the set $\{1,2\}$, i.e., $D=2$ and $\omega=1$, and the responder lies with probability $q=0.1$. Moreover, we set the threshold $\epsilon=0.05$. At the initialization stage, we have $\boldsymbol{\pi}(0)=[0.5,0.5]$. 
In the first round, the query set $\mathcal{S}_1$ is defined as $\{1\}$ based on (\ref{eqn:query set sortPM}) and (\ref{eqn:l star sortPM}), but the responder lies and says $Y_1=0$. Then, we update $\boldsymbol{\pi}(1)=[0.1,0.9]$ based on $Y_1$ according to (\ref{eqn:pi k-1 0}). 
In the second round, $\mathcal{S}_2$ is defined as $\{2\}$ based on (\ref{eqn:query set sortPM}) and (\ref{eqn:l star sortPM}), and the responder does not lie and says $Y_2=0$. Then, we update $\boldsymbol{\pi}(2)=[0.5,0.5]$ based on $Y_2$ according to (\ref{eqn:pi k-1 0}).
In the third round, $\mathcal{S}_3$ is defined as $\{1\}$ based on (\ref{eqn:query set sortPM}) and (\ref{eqn:l star sortPM}), and the responder does not lie and says $Y_3=1$. Then, we update $\boldsymbol{\pi}(3)=[0.9,0.1]$ based on $Y_3$ according to (\ref{eqn:pi k-1 1}).
In the forth round, $\mathcal{S}_4$ is defined as $\{1\}$ based on (\ref{eqn:query set sortPM}) and (\ref{eqn:l star sortPM}), and the responder does not lie and says $Y_4=1$. Then, we update $\boldsymbol{\pi}(4)=[81/82,1/82]$ based on $Y_4$ according to (\ref{eqn:pi k-1 1}).
Because $\pi_1^{\downarrow}(4)>1-\epsilon$, the sortPM method converges and $\hat{\omega}=1$. Although the responder lies in the first round of Q\&A, the questioner can exploit the sortPM method to correctly guess $\omega$ through 4 questions.
\end{example}
\subsection{Detecting the Defective IRS Elements}
In this section, we will focus on the design of sortPM-based method to guess $n_{h,{\rm min}}$ in the set $\{1,\dots,N_h\}$, while the same approach can also be utilized to guess $n_{h,{\rm max}}$, $n_{v,{\rm min}}$, and $n_{v,{\rm max}}$. \par
Note that in the noisy 20 questions problem, after the questioner asks whether $\omega\in\mathcal{S}_k$ at the $k$-th Q\&A round, the responder needs to provide the answer $Y_k$. Under our considered IRS diagnosis problem, the challenge is how can we get an answer about whether $n_{h,{\rm min}} \in\mathcal{S}_k$ in each round $k$ based on the over-the-air method. In the following, we first show, given any IRS sub-region, how to determine whether there exists any defective element at this sub-region, just based on the over-the-air measurements. Then, we will describe how to utilize the above result to determine whether $n_{h,{\rm min}}\in\mathcal{S}_k$, i.e., $Y_k=1$, or $n_{h,{\rm min}}\notin\mathcal{S}_k$, i.e., $Y_k=0$.\par
\subsubsection{Detecting Existence of Defective Elements on a Sub-Region on IRS}
In the $k$-th round of Q\&A, we set the particular sub-region on the IRS, in which the set of all the IRS elements is defined as
\begin{align}\label{eqn:query region}
	\mathcal{U}_{k_t}=\{(n_h,n_v)|n_h\in\mathcal{S}_{k_t},n_v\in\mathcal{N}_v\},
\end{align}
where $\mathcal{S}_{k_t}\subseteq\mathcal{N}_h$ is the set of columns that we are interested for the $t$-th time slot of round $k$. Given any sub-region defined by $\mathcal{U}_{k_t}$, our objective is to detect whether there exists any defective IRS element or not. Depending on the location of the defective IRS elements defined in (\ref{eqn:defective element set}), there are two cases that may occur.\par
\noindent {\bf Case A:} There exist no defective IRS elements in the sub-region defined by $\mathcal{U}_{k_t}$.\par
\noindent {\bf Case B:} There exist some defective IRS elements in the sub-region defined by $\mathcal{U}_{k_t}$.\par
In the following, we will propose an efficient detection scheme such that given any sub-region defined by $\mathcal{U}_{k_t}$, we can determine whether Case A or Case B is true. \par 
Under our proposed scheme, there is an initialization stage which consists of two time slots, denoted by time slots $0_-$ and $0_+$ for convenience. In time slot $0_-$ and $0_+$, we respectively set a common desired phase shift for all the IRS elements as
\begin{align}
    &\phi_{n_h,n_v,0_-}=\bar{\phi}_{0_-},~\forall n_h, n_v, \label{eqn:phase shift 0-} \\
    &\phi_{n_h,n_v,0_+}=\bar{\phi}_{0_+}\neq\bar{\phi}_{0_-}, ~\forall n_h, n_v. \label{eqn:phase shift 0+}
\end{align}
Define
\begin{align}
    &\boldsymbol{g}_{\rm e}=\sum_{(n_h,n_v)\in\mathcal{E}}e^{j\beta_{n_h,n_v}}\boldsymbol{g}_{n_h,n_v}, \label{eqn:ge}\\
    &\boldsymbol{g}_{\rm w}=\sum_{(n_h,n_v)\in\mathcal{W}}\boldsymbol{g}_{n_h,n_v}.\label{eqn:gw}
\end{align}
According to (\ref{eqn:received signal general}), the received signals in time slots $0_-$ and $0_+$ are
\begin{align}
    & \boldsymbol{y}_{0_-}=\boldsymbol{h}x_{0_-}+\boldsymbol{g}_{\rm e}x_{0_-}+\boldsymbol{g}_{\rm w}e^{j\bar{\phi}_{0_-}}x_{0_-}+\boldsymbol{z}_{0_-}, \label{eqn:signal 0-}\\
    & \boldsymbol{y}_{0_+}=\boldsymbol{h}x_{0_+}+\boldsymbol{g}_{\rm e}x_{0_+}+\boldsymbol{g}_{\rm w}e^{j\bar{\phi}_{0_+}}x_{0_+}+\boldsymbol{z}_{0_+}. \label{eqn:signal 0+}
\end{align}
Because $\boldsymbol{z}_{0_-}$, $\boldsymbol{z}_{0_+}\sim\mathcal{CN}(\boldsymbol{0},\sigma^2\boldsymbol{I})$ are independent noise, we have
\begin{align}
    &p(\boldsymbol{y}_{0_-},\boldsymbol{y}_{0_+}|\boldsymbol{g}_{\rm e},\boldsymbol{g}_{\rm w}) \nonumber \\
    = &\frac{1}{(\pi\sigma^2)^{2M}}e^{-\frac{\sum_{i\in\{0_-,0_+\}}\Vert \boldsymbol{y}_{i}-\boldsymbol{h}x_{i}-\boldsymbol{g}_{\rm e}x_{i}-\boldsymbol{g}_{\rm w}e^{j\bar{\phi}_{i}}x_{i}\Vert^2 }{\sigma^2}}.
\end{align}
Then, the ML estimators of $\boldsymbol{g}_{\rm e}$ and $\boldsymbol{g}_{\rm w}$ that maximize the above conditional probability are given as
\begin{align}
& \bar{\boldsymbol{g}}_{\rm e}=\frac{e^{j\bar{\phi}_{0_+}}x_{0_+}(\boldsymbol{y}_{0_-}-\boldsymbol{h}x_{0_-})-e^{j\bar{\phi}_{0_-}}x_{0_-}(\boldsymbol{y}_{0_+}-\boldsymbol{h}x_{0_+})}
{x_{0_-}x_{0_+}(e^{j\bar{\phi}_{0_+}}-e^{j\bar{\phi}_{0_-}})}, \label{eqn:ML of ge} \\
& \bar{\boldsymbol{g}}_{{\rm w}}=\frac{x_{0_-}(\boldsymbol{y}_{0_+}-\boldsymbol{h}x_{0_+})-
x_{0_+}(\boldsymbol{y}_{0_-}-\boldsymbol{h}x_{0_-})}{x_{0_-}x_{0_+}(e^{j\bar{\phi}_{0_+}}-e^{j\bar{\phi}_{0_-}})}. \label{eqn:ML of gw}
\end{align}

After the above initialization stage, given any sub-region $\mathcal{U}_{k_t}$, we respectively set a common desired phase shift for all the IRS elements in this region and one for all the IRS elements out of this region as
\begin{align}
&\phi_{n_h,n_v,k_t}=\bar{\phi}_{k_t,{\rm in}},~{\rm if} ~ n_h\in\mathcal{S}_{k_t}, \label{eqn:phi t in} \\ 
&\phi_{n_h,n_v,k_t}=\bar{\phi}_{k_t,{\rm out}},~{\rm if} ~ n_h\notin \mathcal{S}_{k_t}, ~t\geq 1, \label{eqn:phi t out}
\end{align}
where $\bar{\phi}_{k_t,{\rm in}}\neq\bar{\phi}_{k_t,{\rm out}}$. According to (\ref{eqn:received signal general}) and (\ref{eqn:ge}), the received signal in the $t$-th time slot of round $k$ is given as
\begin{align}\label{eqn:received signal sortPM}
    \boldsymbol{y}_{k_t}=&\boldsymbol{h}x_{k_t}+\boldsymbol{g}_{\rm e}x_{k_t}+\sum_{(n_h,n_v)\in\mathcal{W}_{{k_t},{\rm in}}}\boldsymbol{g}_{n_h,n_v}e^{\bar{\phi}_{{k_t},{\rm in}}}x_{k_t}\nonumber \\ &+\sum_{(n_h,n_v)\in\mathcal{W}_{{k_t},{\rm out}}}\boldsymbol{g}_{n_h,n_v}e^{\bar{\phi}_{{k_t},{\rm out}}}x_{k_t}+\boldsymbol{z}_{k_t},~t\geq 1, 
\end{align}
where
\begin{align}
    &\mathcal{W}_{{k_t},{\rm in}}=\{(n_h,n_v)|(n_h,n_v)\in\mathcal{W},n_h\in\mathcal{S}_{k_t}\}, \\
    &\mathcal{W}_{{k_t},{\rm out}}=\{(n_h,n_v)|(n_h,n_v)\in\mathcal{W},n_h\notin\mathcal{S}_{k_t}\},
\end{align}
denote the sets of the indices of all the normal IRS elements in and out of the sub-region defined by $\mathcal{U}_{k_t}$, respectively. Based on the signal received in the $t$-th time slot of round $k$ as well as the ML estimators made based on the signals received at time slots $0_-$ and $0_+$, our goal is to decide which of Case A and Case B is true in the $t$-th time slot of round $k$.\par
Suppose Case A is true, which indicates that all the IRS elements in $\mathcal{U}_{k_t}$ are in the normal state, i.e., $\mathcal{W}_{{k_t},{\rm in}}=\mathcal{U}_{k_t}$. In this case, define
\begin{align}
    &\boldsymbol{g}_{{k_t},{\rm in}}^{\rm C1}=\sum_{n_h\in\mathcal{S}_{k_t}}\sum_{n_v\in\mathcal{N}_v}\boldsymbol{g}_{n_h,n_v},\label{eqn:gtin C1} \\
    &\boldsymbol{g}_{{k_t},{\rm out}}^{\rm C1}=\bar{\boldsymbol{g}}_{\rm w}-\boldsymbol{g}_{{k_t},{\rm in}}^{\rm C1},\label{eqn:gtout C1}
\end{align}
as the estimates of $\sum_{(n_h,n_v)\in\mathcal{W}_{{k_t},{\rm in}}}\boldsymbol{g}_{n_h,n_v}$ and $\sum_{(n_h,n_v)\in\mathcal{W}_{{k_t},{\rm out}}}\boldsymbol{g}_{n_h,n_v}$ in (\ref{eqn:received signal sortPM}) under Case A. Because $\boldsymbol{z}_{k_t}\sim\mathcal{CN}(\boldsymbol{0},\sigma^2\boldsymbol{I})$, the probability to receive $\boldsymbol{y}_{k_t}$ in the $t$-th time slot of round $k$ under Case A is
\begin{align}\label{eqn:probability sortPM}
    &p(\boldsymbol{y}_{k_t}|{\rm Case~A}) \nonumber \\
    =&\frac{1}{(\pi\sigma^2)^M}e^{-\frac{\Vert \boldsymbol{y}_{k_t}-\boldsymbol{h}x_{k_t}-\bar{\boldsymbol{g}}_{\rm e}x_{k_t}-\boldsymbol{g}_{{k_t},{\rm in}}^{\rm C1}e^{j\bar{\phi}_{{k_t},{\rm in}}}x_{k_t}-\boldsymbol{g}_{{k_t},{\rm out}}^{\rm C1}e^{j\bar{\phi}_{{k_t},{\rm out}}}x_{k_t} \Vert^2}{\sigma^2}}.
\end{align}
Note that based on the received signals at time slots $0_-$ and $0_+$ given in (\ref{eqn:signal 0-}) and (\ref{eqn:signal 0+}), we estimate $2M$ variables in $\boldsymbol{g}_{\rm e}$ and $\boldsymbol{g}_{\rm w}$ via $2M$ equations. Therefore, it is expected that the ML estimators given in (\ref{eqn:ML of ge}) and (\ref{eqn:ML of gw}) are very accurate. If Case A is true, then $\boldsymbol{g}_{{k_t},{\rm in}}^{\rm C1}$ and $\boldsymbol{g}_{{k_t},{\rm out}}^{\rm C1}$ given in (\ref{eqn:gtin C1}) and (\ref{eqn:gtout C1}) are also very accurate estimators of $\sum_{(n_h,n_v)\in\mathcal{W}_{{k_t},{\rm in}}}\boldsymbol{g}_{n_h,n_v}$ and $\sum_{(n_h,n_v)\in\mathcal{W}_{{k_t},{\rm out}}}\boldsymbol{g}_{n_h,n_v}$. To summarize, it is very likely that the conditional probability given in (\ref{eqn:probability sortPM}) will be very high if Case A is true. Otherwise, if Case A is not true, i.e., Case B is true, then $\boldsymbol{g}_{{k_t},{\rm in}}^{\rm C1}$ and $\boldsymbol{g}_{{k_t},{\rm out}}^{\rm C1}$ given in (\ref{eqn:gtin C1}) and (\ref{eqn:gtout C1}) are poor estimators. Therefore, a very large noise power of $\boldsymbol{z}_{k_t}$ can make the equation in (\ref{eqn:received signal sortPM}) hold given the above poor estimators. To summarize, it is very likely that the conditional probability given in (\ref{eqn:probability sortPM}) will be very low if Case A is not true. To summarize, we claim that Case A is true if and only if
\begin{align}\label{eqn:Case A true}
	p(\boldsymbol{y}_{k_t}|{\rm Case~A})>\psi,
\end{align}
where $\psi$ is a pre-determined threshold. \par 
Moreover, we claim that Case B is true if and only if
\begin{align}\label{eqn:Case B true}
	p(\boldsymbol{y}_{k_t}|{\rm Case~A})\leq\psi.
\end{align}
\subsubsection{How to Get $Y_k$'s}
In the previous subsection, we showed how to determine whether there exists any defective IRS element within any region $\mathcal{U}_{k_t}$ defined in (\ref{eqn:query region}). Next, we will show that given any $\mathcal{S}_k$ in the $k$-th round, we need at most two time slots to detect two sub-regions on the IRS so as to determine whether $n_{h,{\rm min}}$ is in the query set $\mathcal{S}_k$ or not.\par
\begin{figure}
	\centering
	\includegraphics[scale=0.48]{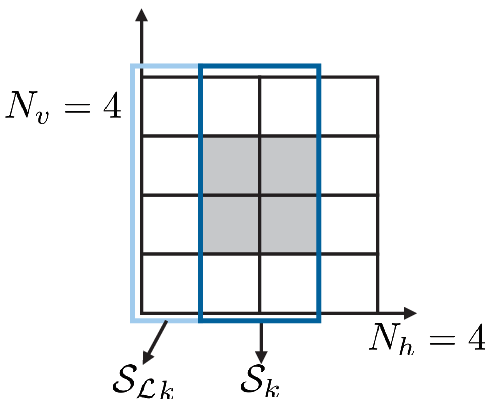}
	\caption{An IRS with $N_h=4$ columns and $N_v=4$ rows: $\mathcal{S}_k=\{2,3\}$, and $\mathcal{U}_k=\{(n_h,n_v)|n_h\in\{2,3\},n_v\in\mathcal{N}_v\}$; $\mathcal{S_L}_k=\{1\}$, and $\mathcal{U_L}_k=\{(n_h,n_v)|n_h=1,n_v\in\mathcal{N}_v\}$.}
	\label{fig3}
\end{figure}
Let $\mathcal{S_L}_k$ denote the set of columns that are on the left hand side of $\mathcal{S}_k$, as shown in Fig. \ref{fig3}. Similar to (\ref{eqn:query region}), the sub-regions in which $\mathcal{S}_k$ and $\mathcal{S_L}_k$ are the set of columns are respectively given as
\begin{align}
	\mathcal{U}_{k}&=\{(n_h,n_v)|n_h\in\mathcal{S}_k,n_v\in\mathcal{N}_v\}, \\
	\mathcal{U_L}_{k}&=\{(n_h,n_v)|n_h\in\mathcal{S_L}_k,n_v\in\mathcal{N}_v\}.
\end{align}

Suppose $n_{h,{\rm min}}\in\mathcal{S}_k$, i.e., there exist defective elements in the sub-region $\mathcal{U}_{k}$ and there does not exist any defective element in the sub-region $\mathcal{U_L}_{k}$. In this case, the over-the-air measurements should also indicate that Case B is true for $\mathcal{U}_k$, and Case A is true for $\mathcal{U_L}_k$. Therefore, in the first time slot of round $k$, we perform diagnosis for the sub-region $\mathcal{U}_{k}$, i.e., $\mathcal{U}_{k_1}=\mathcal{U}_{k}$. If Case A is true for $\mathcal{U}_{k_1}$, we directly claim that $Y_k=0$. If Case B is true for $\mathcal{U}_{k_1}$, then in the second time slot of round $k$, we perform diagnosis for the sub-region $\mathcal{U_L}_{k}$, i.e., $\mathcal{U}_{k_2}=\mathcal{U_L}_{k}$. If Case A is true for $\mathcal{U}_{k_2}$, we claim that $Y_k=1$ in two time slots. If Case B is true for $\mathcal{U}_{k_2}$, we claim that $Y_k=0$ in two time slots. To summarize, the answer $Y_k$ in the $k$-th round of Q\&A is given as
\begin{align}\label{eqn:Y_k sortPM}
	Y_{k}=\left\{\begin{array}{ll} 
		1, &
		\begin{array}{ll}
			{\rm if}&p(\boldsymbol{y}_{k_1}|{\rm Case~A})\leq\psi, \\
			&p(\boldsymbol{y}_{k_2}|{\rm Case~A})>\psi,
		\end{array} \\
		0, &{\rm ~otherwise}, \end{array}\right.\forall k\geq 1.
\end{align}
\subsubsection{Our Algorithm}
After showing how to determine whether $n_{h,{\rm min}}$ is in $\mathcal{S}_k$, we propose a novel sortPM-based method to estimate $n_{h,{\rm min}}$. In time slots $0_-$ and $0_+$, we set desired IRS phase shifts according to (\ref{eqn:phase shift 0-}) and (\ref{eqn:phase shift 0+}), keep a record of the received signals shown in (\ref{eqn:signal 0-}) and (\ref{eqn:signal 0+}), and make ML estimations of $\boldsymbol{g}_{\rm e}$ and $\boldsymbol{g}_{\rm w}$ based on (\ref{eqn:ML of ge}) and (\ref{eqn:ML of gw}). After the above initialization stage, based on the sortPM method in Algorithm 1, our method for estimating $n_{h,{\rm min}}$ is summarized in Algorithm \ref{alg:sortPM 1}.\footnote{In our considered IRS diagnosis problem, we assume that all the elements in $\mathcal{S}_k$ are consecutive integers. If this does not hold at some round $k$, we still sort all the posterior probabilities and $\mathcal{S}_k$ contains the single index with the largest posterior probability, i.e., $\gamma_{1}(k-1)$.}\par
\begin{algorithm}
   \caption{SortPM-based Method for Estimating $n_{h,{\rm min}}$}
   \label{alg:sortPM 1}
  {\bf Input}: a positive threshold $\epsilon$ close to 0, $N_h$, $N_v$, and cascade channel $\boldsymbol{g}_{n_h,n_v}$, $\forall n_h\in\mathcal{N}_h, n_v\in\mathcal{N}_v$\;
  {\bf Initialization}: $\pi_{d}(0)=1/N_h$, $\forall d\in\mathcal{N}_h$\;
  Perform the initialization stage and get the ML estimators of $\boldsymbol{g}_{\rm e}$ and $\boldsymbol{g}_{\rm w}$ based on (\ref{eqn:ML of ge}) and (\ref{eqn:ML of gw})\;
  \For{$k=1,2,\dots$}
  {
  	Design the query set $\mathcal{S}_k$ based on (\ref{eqn:query set sortPM}) and (\ref{eqn:l star sortPM})\;
  	Perform Diagnosis for $\mathcal{U}_k$ and $\mathcal{U_L}_k$\;
  	\eIf{\rm (\ref{eqn:Case A true}) is true for $\mathcal{U}_k$}
  	{$Y_k=0$\;}
  	{\eIf{\rm (\ref{eqn:Case B true}) is true for $\mathcal{U_L}_k$}
  		{$Y_k=0$\;}{$Y_k=1$\;}
  	}
  	Update $\pi_d(k),~\forall d=1,\dots,N_h,$ based on ${Y}_{k}$, according to (\ref{eqn:pi k-1 1}) and (\ref{eqn:pi k-1 0})\;
  	\If{$\pi_{1}^{\downarrow}(k)\geq 1-\epsilon$}
  	{break\;} 
  }	
  The estimation of $n_{h,{\rm min}}$: $\hat{n}_{h,{\rm min}}=\gamma_{1}(k)$\;
  {\bf Output}: $\hat{n}_{h,{\rm min}}$\;
\end{algorithm}
\section{Bisection Method for Localizing Defective IRS Elements}
In this paper, we assume that the IRS is in the anechoic chamber. Therefore, the SNR for over-the-air diagnosis is high. In this case, the answers $Y_k$'s obtained in Section III are very accurate. This motivates us to design a diagnosis method based on the noiseless 20 questions technique in this section, which is of lower complexity than the method proposed in the previous section.\par
The bisection method belongs to the class of the cutting-plane methods, and is widely used to solve the one-dimensional optimization problem. We will show that the bisection method can also be exploited to localize the defective IRS elements under the over-the-air scheme. For convenience, we will focus on the design of the bisection-based method to estimate $n_{h,{\rm min}}$ and $n_{h,{\rm max}}$ in the horizontal dimension, while the same approach can be utilized to estimate $n_{v,{\rm min}}$ and $n_{v,{\rm max}}$ in the vertical dimension.\par
The basic idea of our proposed scheme is as follows. Starting with the interval $[1,N_h]$, we keep cutting off half of the interval that does not contain $n_{h,{\rm min}}$ ($n_{h,{\rm max}}$) until the remaining interval is sufficiently small such that the middle point of this remaining interval can serve as the estimation of $n_{h,{\rm min}}$ ($n_{h,{\rm max}}$). In the following, we first show, given any boundary defined by $c_h=\bar{n}$, how to determine whether we should cut off the interval on the left hand side of this boundary with $n_h<\bar{n}$ or the interval on the right hand side of this boundary with $n_h>\bar{n}$, just based on the over-the-air measurements. Then, we will introduce our proposed bisection method to iteratively update $\bar{n}$ and estimate $n_{h,{\rm min}}$ and $n_{h,{\rm max}}$.\par
\subsection{Determining the Cutting-Plane}
Given any boundary defined by $c_h=\bar{n}$, our goal is to determine whether $n_{h,{\rm min}}$ ($n_{h,{\rm max}}$) is on the left hand side or on the right hand side of this boundary. Depending on the location of the defective IRS elements defined in (\ref{eqn:defective element set}), there are three cases that may occur.\par
\begin{figure}
    \centering
    \includegraphics[width=0.9\linewidth]{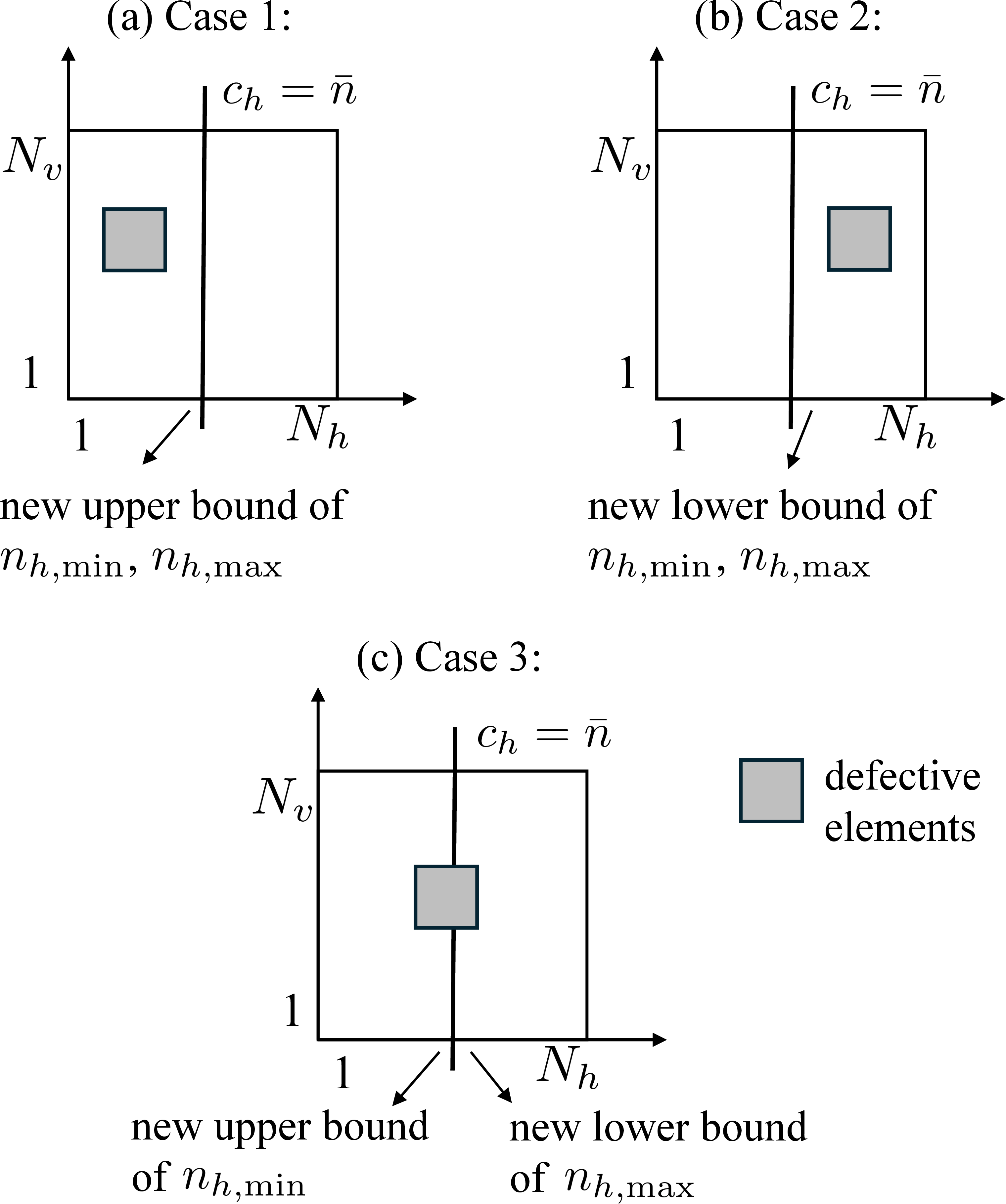}
    \caption{ Three cases of the location of defective elements related to the
boundary and the corresponding update of the lower and upper bounds.}
    \label{fig4}
\end{figure}

\noindent {\bf Case 1:} All the defective IRS elements defined in (\ref{eqn:defective element set}) are on the left hand side of the boundary defined by $c_h=\bar{n}$, i.e., $n_{h,{\rm min}}\leq n_{h,{\rm max}}<\bar{n}$, as shown in Fig. \ref{fig4} (a).\par
\noindent {\bf Case 2:} All the defective IRS elements defined in (\ref{eqn:defective element set}) are on the right hand side of the boundary defined by $c_h=\bar{n}$, i.e., $n_{h,{\rm max}}\geq n_{h,{\rm min}}>\bar{n}$, as shown in Fig. \ref{fig4} (b).\par
\noindent {\bf Case 3:} The defective IRS elements defined in (\ref{eqn:defective element set}) are on both the left hand side and the right hand side of the boundary defined by $c_h=\bar{n}$, i.e., $n_{h,{\rm min}}<\bar{n}$ and $n_{h,{\rm max}}>\bar{n}$, as shown in Fig. \ref{fig4} (c).\par

In the following, we propose an efficient detection scheme such that given any boundary defined by $c_h=\bar{n}$, we can determine which of the above three cases is true.\par
After the initialization stage defined in (\ref{eqn:signal 0-})-(\ref{eqn:ML of gw}), we make some boundary defined by $c_h=\bar{n}_t$ at each time slot $t\geq 1$, which can divide the whole IRS into two parts. At time slot $t$, we respectively set a common desired phase shift for all the IRS elements on the left hand side of the boundary and one for all the IRS elements on the right hand side of the boundary as
\begin{align}
    &\phi_{n_h,n_v,t}=\bar{\phi}_{t,{\rm l}},~ {\rm if}~ n_h\in[1,\lfloor\bar{n}_t\rfloor], \label{eqn:phase shift left}\\
    &\phi_{n_h,n_v,t}=\bar{\phi}_{t,{\rm r}},~ {\rm if}~ n_h\in[\lceil\bar{n}_t\rceil,N_h], ~ t\geq 1, \label{eqn:phase shift right}
\end{align}
where $\bar{\phi}_{t,{\rm l}}\neq\bar{\phi}_{t,{\rm r}}$. According to (\ref{eqn:received signal general}) and (\ref{eqn:ge}), the received signal at time slot $t\geq 1$ is
\begin{align}
    \boldsymbol{y}_{t}=&\boldsymbol{h}x_{t}+\boldsymbol{g}_{\rm e}x_{t}+\sum_{(n_h,n_v)\in\mathcal{W}_{t,{\rm l}}}\boldsymbol{g}_{n_h,n_v}e^{j\bar{\phi}_{t,{\rm l}}}x_t \nonumber \\
    &+\sum_{(n_h,n_v)\in\mathcal{W}_{t,{\rm r}}}\boldsymbol{g}_{n_h,n_v}e^{j\bar{\phi}_{t,{\rm r}}}x_t+\boldsymbol{z}_{t}, ~ t\geq 1, \label{eqn:received signal}
\end{align}
where
\begin{align}
    &\mathcal{W}_{t,{\rm l}}=\{(n_h,n_v)|(n_h,n_v)\in\mathcal{W},n_h<\bar{n}_t\}, \label{left normal set}\\
    &\mathcal{W}_{t,{\rm r}}=\{(n_h,n_v)|(n_h,n_v)\in\mathcal{W},n_h>\bar{n}_t\}, \label{right normal set}
\end{align}
denote the sets of the normal IRS elements that are on the left hand side and on the right hand side of the boundary defined by $c_h=\bar{n}_t$, respectively.  Based on the signal received in time slot $t$ as well as the ML estimators made based on the signals received at time slots $0_-$ and $0_+$, our goal is to decide which of Case 1, Case 2, and Case 3 is true at time slot $t$.\par
Suppose Case 1 is true. This indicates that all the IRS elements on the right hand side of the boundary $c_h=\bar{n}_t$ are in the normal state, i.e., $\mathcal{W}_{t,{\rm r}}=\{(n_h,n_v)|n_h>\bar{n}_t\}$. In this case, define
\begin{align}
    &\boldsymbol{g}_{t,{\rm r}}^{\rm C1}=\sum_{n_h>\bar{n}_t}\sum_{n_v\in\mathcal{N}_v}\boldsymbol{g}_{n_h,n_v}, \label{eqn:gtl C1} \\
    &\boldsymbol{g}_{t,{\rm l}}^{\rm C1}=\bar{\boldsymbol{g}}_{\rm w}-\boldsymbol{g}_{t,{\rm r}}^{\rm C1}, \label{eqn:gtr C1}
\end{align}
as the estimates of $\sum_{(n_h,n_v)\in\mathcal{W}_{t,{\rm r}}}\boldsymbol{g}_{n_h,n_v}$ and $\sum_{(n_h,n_v)\in\mathcal{W}_{t,{\rm l}}}\boldsymbol{g}_{n_h,n_v}$ in (\ref{eqn:received signal}) under Case 1. Because $\boldsymbol{z}_t\sim\mathcal{CN}(\boldsymbol{0},\sigma^2\boldsymbol{I})$, the probability to receive $\boldsymbol{y}_t$ at time slot $t$ given the estimates $\bar{\boldsymbol{g}}_{\rm e}$, $\boldsymbol{g}_{t,{\rm l}}^{\rm C1}$, and $\boldsymbol{g}_{t,{\rm r}}^{\rm C1}$ is
\begin{align}
    &p(\boldsymbol{y}_{t}|\bar{\boldsymbol{g}}_{\rm e},\boldsymbol{g}_{t,{\rm l}}^{\rm C1},\boldsymbol{g}_{t,{\rm r}}^{\rm C1}) \nonumber \\
    = &\frac{1}{(\pi\sigma^2)^{M}}e^{-\frac{\Vert \boldsymbol{y}_{t}-\boldsymbol{h}x_{t}-\bar{\boldsymbol{g}}_{\rm e}x_{t}-\boldsymbol{g}_{t,{\rm l}}^{\rm C1}e^{j\bar{\phi}_{t,{\rm l}}}x_{t}-\boldsymbol{g}_{t,{\rm r}}^{\rm C1}e^{j\bar{\phi}_{t,{\rm r}}}x_{t}\Vert^2 }{\sigma^2}}. \label{eqn:probability Case 1}
\end{align}
Similar to the detection method proposed in the previous section, we claim that Case 1 is true if and only if
\begin{align}
    p(\boldsymbol{y}_{t}|\bar{\boldsymbol{g}}_{\rm e},\boldsymbol{g}_{t,{\rm l}}^{\rm C1},\boldsymbol{g}_{t,{\rm r}}^{\rm C1})>\bar{p}_1 \label{eqn:threshold 1},
\end{align}
where $\bar{p}_1$ is a pre-determined threshold.\par
Next, if Case 2 is true, all the IRS elements on the left hand side of the boundary $c_h=\bar{n}_t$ are in the normal state, i.e., $\mathcal{W}_{t,{\rm l}}=\{(n_h,n_v)|n_h<\bar{n}_t\}$. Define
\begin{align}
    &\boldsymbol{g}_{t,{\rm l}}^{\rm C2}=\sum_{n_h<\bar{n}_t}\sum_{n_v\in\mathcal{N}_v}\boldsymbol{g}_{n_h,n_v}, \label{eqn: gtl C2} \\
    &\boldsymbol{g}_{t,{\rm r}}^{\rm C2}=\bar{\boldsymbol{g}}_{\rm w}-\boldsymbol{g}_{t,{\rm l}}^{\rm C2}, \label{eqn:gtr C2}
\end{align}
as the estimations of $\sum_{(n_h,n_v)\in\mathcal{W}_{t,{\rm l}}}\boldsymbol{g}_{n_h,n_v}$ and $\sum_{(n_h,n_v)\in\mathcal{W}_{t,{\rm r}}}\boldsymbol{g}_{n_h,n_v}$ in (\ref{eqn:received signal}) under Case 2. Similar to (\ref{eqn:probability Case 1}), the probability to receive $\boldsymbol{y}_t$ at time slot $t$ given the estimations $\bar{\boldsymbol{g}}_{\rm e}$, $\boldsymbol{g}_{t,{\rm l}}^{\rm C2}$, and $\boldsymbol{g}_{t,{\rm r}}^{\rm C2}$ is
\begin{align}
    &p(\boldsymbol{y}_{t}|\bar{\boldsymbol{g}}_{\rm e},\boldsymbol{g}_{t,{\rm l}}^{\rm C2},\boldsymbol{g}_{t,{\rm r}}^{\rm C2}) \nonumber \\
    = &\frac{1}{(\pi\sigma^2)^{M}}e^{-\frac{\Vert \boldsymbol{y}_{t}-\boldsymbol{h}x_{t}-\bar{\boldsymbol{g}}_{\rm e}x_{t}-\boldsymbol{g}_{t,{\rm l}}^{\rm C2}e^{j\bar{\phi}_{t,{\rm l}}}x_{t}-\boldsymbol{g}_{t,{\rm r}}^{\rm C2}e^{j\bar{\phi}_{t,{\rm r}}}x_{t}\Vert^2 }{\sigma^2}}. \label{eqn:probability Case 2}
\end{align}
We claim that Case 2 is true if and only if
\begin{align}\label{eqn:threshold 2}
    p(\boldsymbol{y}_{t}|\bar{\boldsymbol{g}}_{\rm e},\boldsymbol{g}_{t,{\rm l}}^{\rm C2},\boldsymbol{g}_{t,{\rm r}}^{\rm C2})>\bar{p}_2,
\end{align}
where $\bar{p}_2$ is a pre-determined threshold.\par
At last, we claim that Case 3 is true if and only if both (\ref{eqn:threshold 1}) and (\ref{eqn:threshold 2}) do not hold,\footnote{If both (\ref{eqn:threshold 1}) and (\ref{eqn:threshold 2}) hold, this indicates that all the IRS elements are detected to be in the normal state, i.e., a diagnosis of the IRS is no longer needed. Because this paper assumes that there exist defective elements on the IRS, the above case is very unlikely to occur if $\bar{p}_1$ and $\bar{p}_2$ are properly selected.} i.e.,
\begin{align}
    &p(\boldsymbol{y}_{t}|\bar{\boldsymbol{g}}_{\rm e},\boldsymbol{g}_{t,{\rm l}}^{\rm C1},\boldsymbol{g}_{t,{\rm r}}^{\rm C1})\leq\bar{p}_1, \label{eqn:threshold 3_1}\\
    &p(\boldsymbol{y}_{t}|\bar{\boldsymbol{g}}_{\rm e},\boldsymbol{g}_{t,{\rm l}}^{\rm C2},\boldsymbol{g}_{t,{\rm r}}^{\rm C2})\leq\bar{p}_2. \label{eqn:threshold 3_2}
\end{align}
\subsection{Three-Phase Bisection Method}
After showing how to make the cutting plane, in the following, we propose a novel three-phase bisection method to estimate $n_{h,{\rm min}}$ and $n_{h,{\rm max}}$ in the horizontal dimension, based on the over-the-air measurements.\par
Specifically, define $n_{h,{\rm min}}^{{\rm lb},0}=1$ ($n_{h,{\rm max}}^{{\rm lb},0}=1$) and $n_{h,{\rm min}}^{{\rm ub},0}=N_h$ ($n_{h,{\rm max}}^{{\rm ub},0}=N_h$) as the initial lower and upper bounds on $n_{h,{\rm min}}$ ($n_{h,{\rm max}}$). In time slots $0_-$ and $0_+$, we set desired IRS phase shifts according to (\ref{eqn:phase shift 0-}) and (\ref{eqn:phase shift 0+}), keep a record of the received signals shown in (\ref{eqn:signal 0-}) and (\ref{eqn:signal 0+}), and make ML estimations of $\boldsymbol{g}_{\rm e}$ and $\boldsymbol{g}_{\rm w}$ based on (\ref{eqn:ML of ge}) and (\ref{eqn:ML of gw}). After the above initialization stage, we conduct the $t$-th iteration of the bisection method based on the signal received at time slot $t$ via applying the detection method proposed in the previous subsection. Let $n_{h,{\rm min}}^{{\rm lb},t}$ ($n_{h,{\rm max}}^{{\rm lb},t}$) and $n_{h,{\rm min}}^{{\rm ub},t}$ ($n_{h,{\rm max}}^{{\rm ub},t}$) denote the lower and upper bounds on $n_{h,{\rm min}}$ ($n_{h,{\rm max}}$) obtained after the $t$-th iteration of our bisection method. Depending on which bounds are updated, we divide the whole process of our proposed bisection method into three phases.
\subsubsection{Phase I}
Phase I is the phase where the bounds of $n_{h,{\rm min}}$ and $n_{h,{\rm max}}$ are updated together at one iteration. Specifically, in the first few iterations of the bisection method, $n_{h,{\rm min}}$ and $n_{h,{\rm max}}$ have the same lower and upper bounds. For example, at the initialization stage, we have $n_{h,{\rm min}}^{{\rm lb},0}=n_{h,{\rm max}}^{{\rm lb},0}=1$ and $n_{h,{\rm min}}^{{\rm ub},0}=n_{h,{\rm max}}^{{\rm ub},0}=N_h$. If $n_{h,{\rm min}}^{{\rm lb},t}=n_{h,{\rm max}}^{{\rm lb},t}$ and $n_{h,{\rm min}}^{{\rm ub},t}=n_{h,{\rm max}}^{{\rm ub},t}$ are true at the $t$-th iteration, then at the ($t+1$)-th iteration, the boundary for estimating $n_{h,{\rm min}}$, which is defined by the middle point of the interval $[n_{h,{\rm min}}^{{\rm lb},t}, n_{h,{\rm min}}^{{\rm ub},t}]$, i.e., $n_{h,{\rm min}}^{{\rm md},t+1}=(n_{h,{\rm min}}^{{\rm lb},t}+n_{h,{\rm min}}^{{\rm ub},t})/2$, and the boundary for estimating $n_{h,{\rm max}}$, which is defined by the middle point of the interval $[n_{h,{\rm max}}^{{\rm lb},t}, n_{h,{\rm max}}^{{\rm ub},t}]$, i.e., $n_{h,{\rm max}}^{{\rm md},t+1}=(n_{h,{\rm max}}^{{\rm lb},t}+n_{h,{\rm max}}^{{\rm ub},t})/2$, are the same, i.e., $n_{h,{\rm min}}^{{\rm md},t+1}=n_{h,{\rm max}}^{{\rm md},t+1}$. Therefore, we can simultaneously update the bounds of $n_{h,{\rm min}}$ and $n_{h,{\rm max}}$ at the ($t+1$)-th iteration as follows. At time slot $t+1$, we set the boundary as $c_h=\bar{n}_{t+1}=n_{h,{\rm min}}^{{\rm md},t+1}=n_{h,{\rm max}}^{{\rm md},t+1}$ and the phase shifts of IRS elements based on (\ref{eqn:phase shift left}) and (\ref{eqn:phase shift right}). Based on the signal received at time slot $t+1$, i.e., $\boldsymbol{y}_{t+1}$, if Case 1 is true, i.e., (\ref{eqn:threshold 1}) holds, then we update the new bounds as $n_{h,{\rm min}}^{{\rm lb},t+1}=n_{h,{\rm max}}^{{\rm lb},t+1}=n_{h,{\rm min}}^{{\rm lb},t}$ and $n_{h,{\rm min}}^{{\rm ub},t+1}=n_{h,{\rm max}}^{{\rm ub},t+1}=\lfloor n_{h,{\rm min}}^{{\rm md},t+1}\rfloor$. If Case 2 is true, i.e., (\ref{eqn:threshold 2}) holds, then we update the new bounds as $n_{h,{\rm min}}^{{\rm lb},t+1}=n_{h,{\rm max}}^{{\rm lb},t+1}=\lceil n_{h,{\rm min}}^{{\rm md},t+1}\rceil$ and $n_{h,{\rm min}}^{{\rm ub},t+1}=n_{h,{\rm max}}^{{\rm ub},t+1}=n_{h,{\rm min}}^{{\rm ub},t}$. If Case 3 is true, i.e., both (\ref{eqn:threshold 3_1}) and (\ref{eqn:threshold 3_2}) hold, then we update the new bounds as $n_{h,{\rm min}}^{{\rm lb},t+1}=n_{h,{\rm min}}^{{\rm lb},t}$, $n_{h,{\rm min}}^{{\rm ub},t+1}=\lfloor n_{h,{\rm min}}^{{\rm md},t+1}\rfloor$, $n_{h,{\rm max}}^{{\rm lb},t+1}=\lceil n_{h,{\rm max}}^{{\rm md},t+1}\rceil$, and $n_{h,{\rm max}}^{{\rm ub},t+1}=n_{h,{\rm max}}^{{\rm ub},t}$.\par
Note that if Case 1 or Case 2 is true, we still have $n_{h,{\rm min}}^{{\rm lb},t+1}=n_{h,{\rm max}}^{{\rm lb},t+1}$ and $n_{h,{\rm min}}^{{\rm ub},t+1}=n_{h,{\rm max}}^{{\rm ub},t+1}$ at the $(t+1)$-th iteration. Otherwise, if Case 3 is true, we have $n_{h,{\rm min}}^{{\rm lb},t+1}\neq n_{h,{\rm max}}^{{\rm lb},t+1}$ and $n_{h,{\rm min}}^{{\rm ub},t+1}\neq n_{h,{\rm max}}^{{\rm ub},t+1}$ at the $(t+1)$-th iteration. Define $t^{\rm I}$ as the index of the iteration where Case 3 occurs for the first time. Then, Phase I of our proposed bisection method ends at the $t^{\rm I}$-th iteration, when the boundary passes through the cluster of defective IRS elements for the first time such that the bounds of $n_{h,{\rm min}}$ and $n_{h,{\rm max}}$ are different.\par
One example about the update of the bounds in the three cases of Phase I can be found in Fig. \ref{fig4}.\par
\subsubsection{Phase II}
Phase II starts at time slot $t^{\rm I}+1$ and is the phase where merely the bounds of $n_{h,{\rm min}}$ are updated until $n_{h,{\rm min}}$ is found. At iteration $t+1$ with $t\geq t^{\rm I}$, the new boundary is $c_h=n_{h,{\rm min}}^{{\rm md},t+1}=(n_{h,{\rm min}}^{{\rm lb},t}+n_{h,{\rm min}}^{{\rm ub},t})/2$. Based on the signal received at time slot $t+1$, if Case 2 is true, we set $n_{h,{\rm min}}^{{\rm lb},t+1}=\lceil n_{h,{\rm min}}^{{\rm md},t+1}\rceil$ and $n_{h,{\rm min}}^{{\rm ub},t+1}=n_{h,{\rm min}}^{{\rm ub},t}$. If Case 3 is true, we set $n_{h,{\rm min}}^{{\rm lb},t+1}=n_{h,{\rm min}}^{{\rm lb},t}$ and $n_{h,{\rm min}}^{{\rm ub},t+1}=\lfloor n_{h,{\rm min}}^{{\rm md},t+1}\rfloor$. Define $t^{\rm II}$ as the index of the iteration where $n_{h,{\rm min}}^{{\rm lb},t^{\rm II}}=n_{h,{\rm min}}^{{\rm ub},t^{\rm II}}$. Then, Phase II ends at the $t^{\rm II}$-th iteration and $n_{h,{\rm min}}$ is estimated as $n_{h,{\rm min}}^{{\rm lb},t^{\rm II}}$.
\subsubsection{Phase III}
Phase III starts at time slot $t^{\rm II}+1$ and is the phase where merely the bounds of $n_{h,{\rm max}}$ are updated until $n_{h,{\rm max}}$ is found. At iteration $t+1$ with $t\geq t^{\rm II}$, the new boundary is $c_h=n_{h,{\rm max}}^{{\rm md},t+1}=(n_{h,{\rm max}}^{{\rm lb},t}+n_{h,{\rm max}}^{{\rm ub},t})/2$. Based on the signal received at time slot $t+1$, if Case 1 is true, we set $n_{h,{\rm max}}^{{\rm ub},t+1}=\lfloor n_{h,{\rm max}}^{{\rm md},t+1}\rfloor$ and $n_{h,{\rm max}}^{{\rm lb},t+1}=n_{h,{\rm max}}^{{\rm lb},t}$. If Case 3 is true, we set $n_{h,{\rm max}}^{{\rm lb},t+1}=\lceil n_{h,{\rm max}}^{{\rm md},t+1}\rceil$ and $n_{h,{\rm max}}^{{\rm ub},t+1}=n_{h,{\rm max}}^{{\rm ub},t}$. Define $t^{\rm III}$ as the index of the iteration where $n_{h,{\rm max}}^{{\rm lb},t^{\rm III}}=n_{h,{\rm max}}^{{\rm ub},t^{\rm III}}$. Then, Phase III ends at the $t^{\rm III}$-th iteration and $n_{h,{\rm max}}$ is estimated as $n_{h,{\rm max}}^{{\rm lb},t^{\rm III}}$.\par
\begin{figure}
	\centering
	\subfigure[$t=1$ (Phase I)]{\includegraphics[scale=0.15]{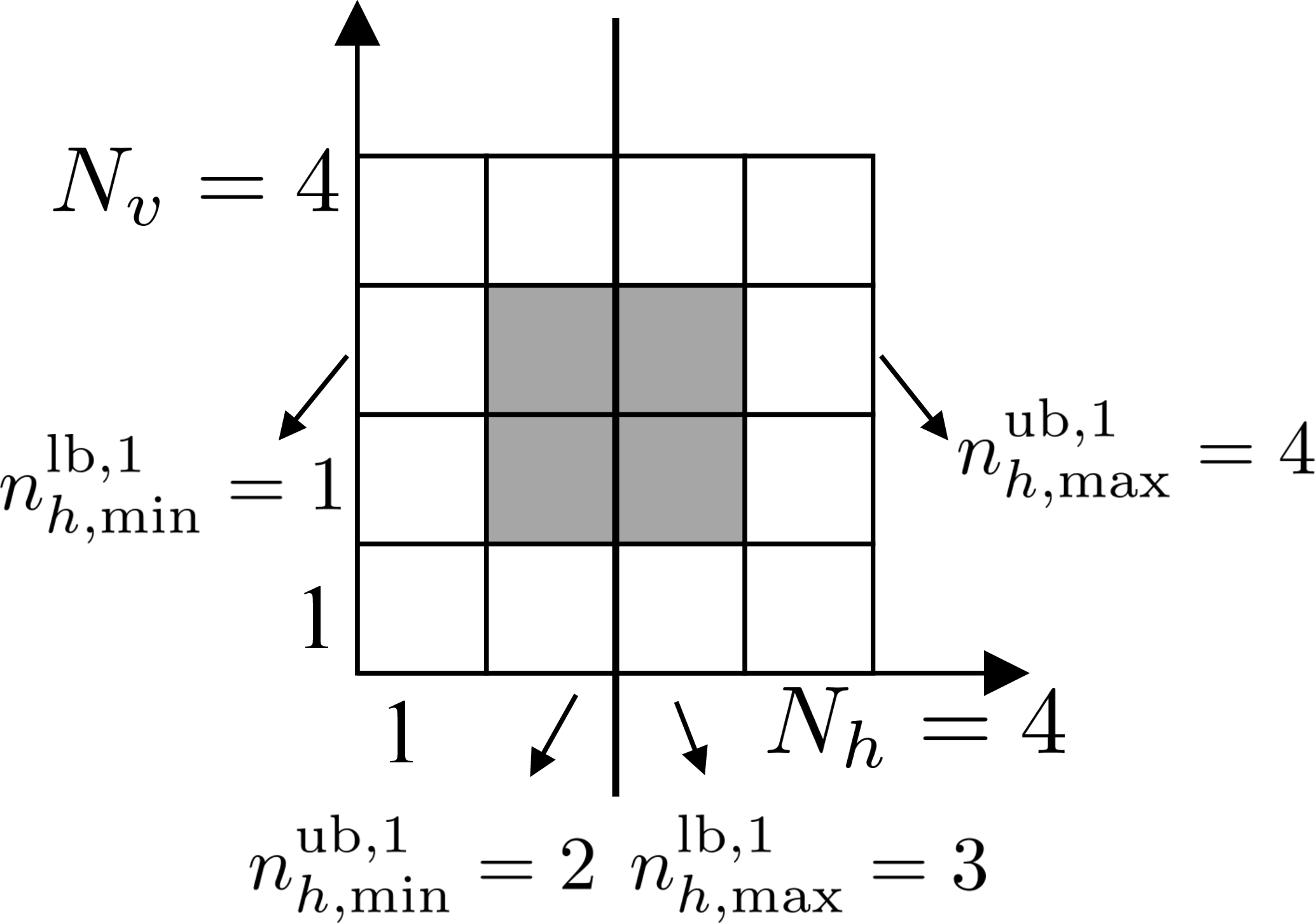}\label{fig5a}}
	\subfigure[$t=2$ (Phase II)]{\includegraphics[scale=0.15]{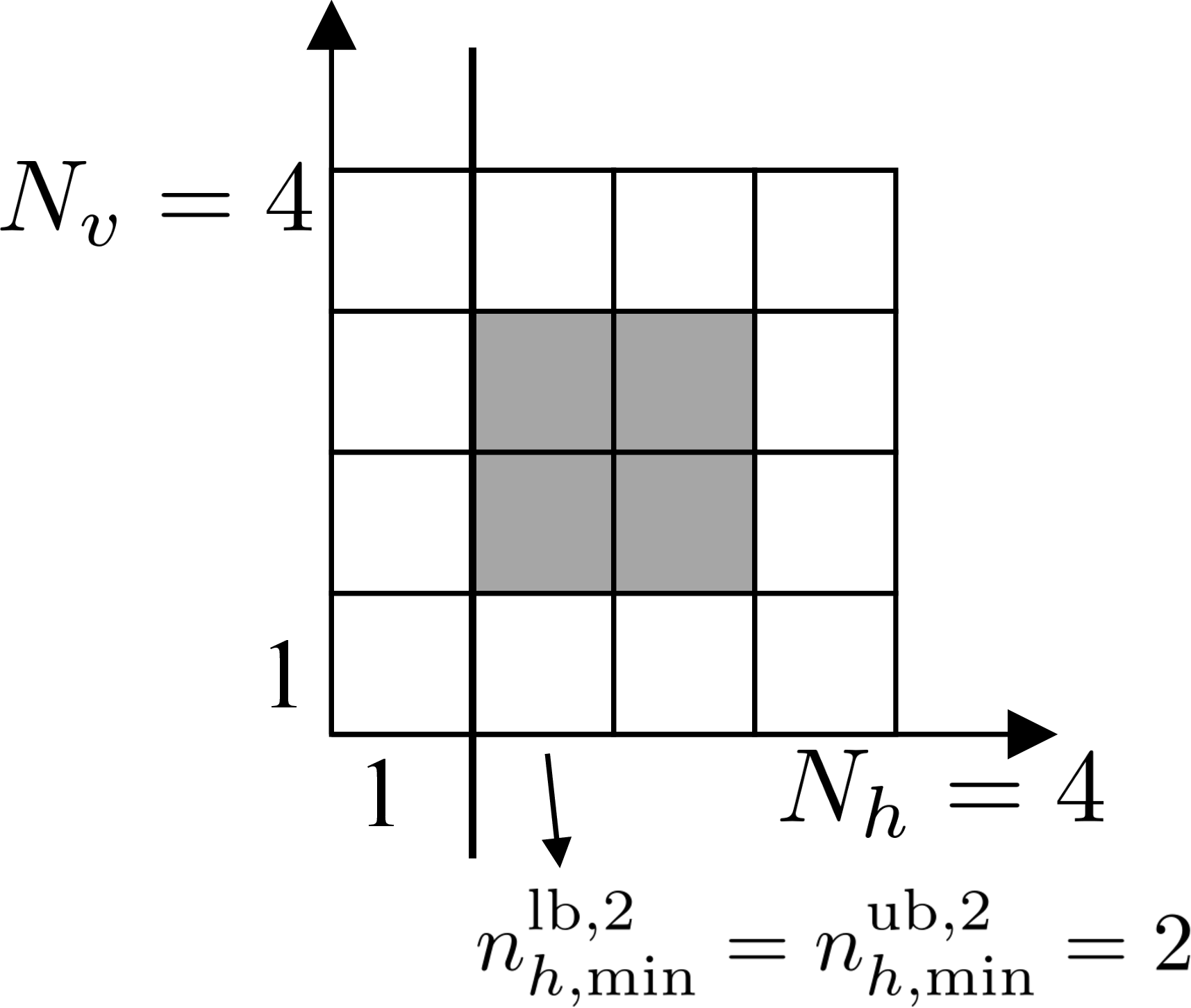}\label{fig5b}}
	\subfigure[$t=3$ (Phase III)]{\includegraphics[scale=0.15]{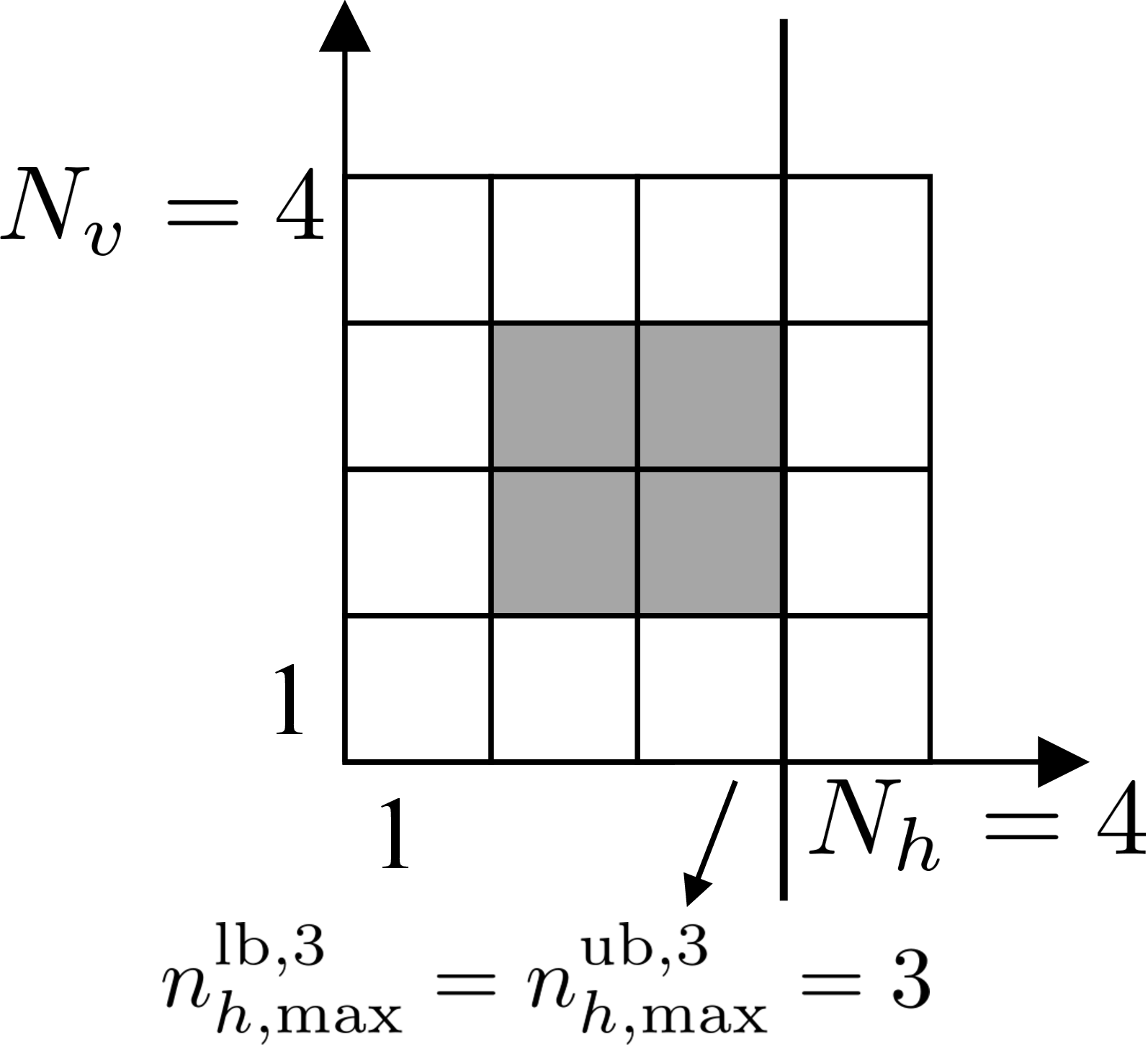}\label{fig5c}}
	\caption{An illustration of our proposed three-phase bisection approach for localizing defective IRS elements.} \vspace{-10pt} \label{fig5}
\end{figure}
\begin{figure}
	\centering
	\subfigure[$k=1$]{\includegraphics[scale=0.34]{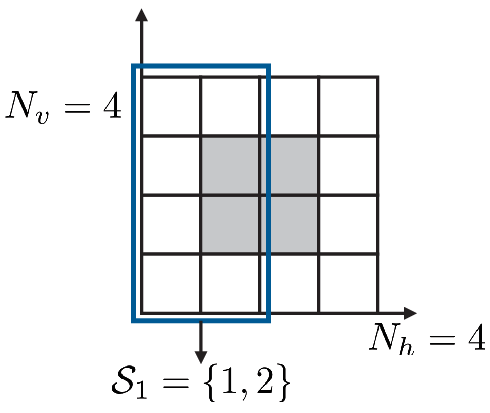}\label{fig6a}}
 \subfigure[$k=2$]{\includegraphics[scale=0.34]{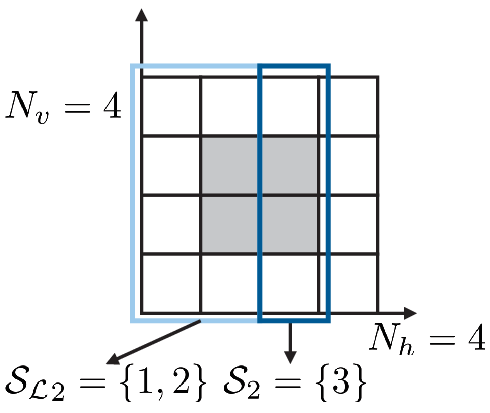}\label{fig6b}}
 \subfigure[$k=3$]{\includegraphics[scale=0.34]{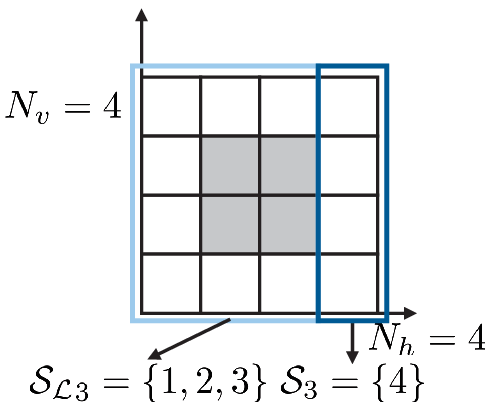}\label{fig6c}}
 \subfigure[$k=4$]{\includegraphics[scale=0.34]{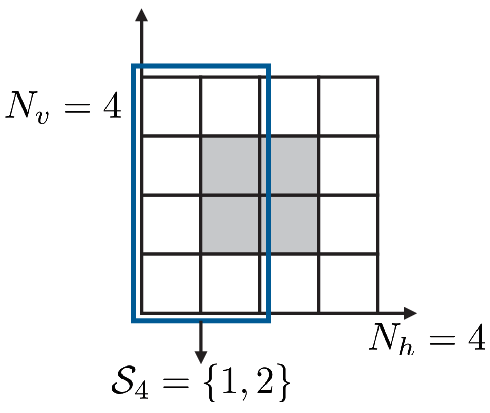}\label{fig6d}}
\subfigure[$k=5$]{\includegraphics[scale=0.34]{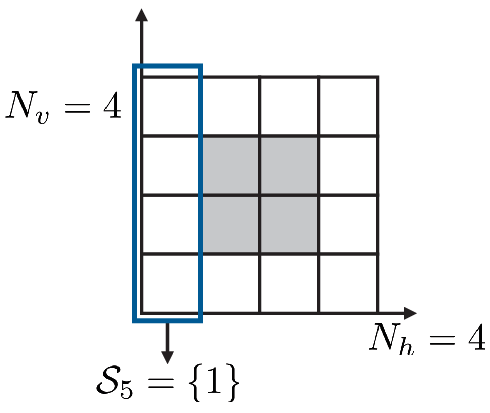}\label{fig6e}}
\subfigure[$k=6$]{\includegraphics[scale=0.34]{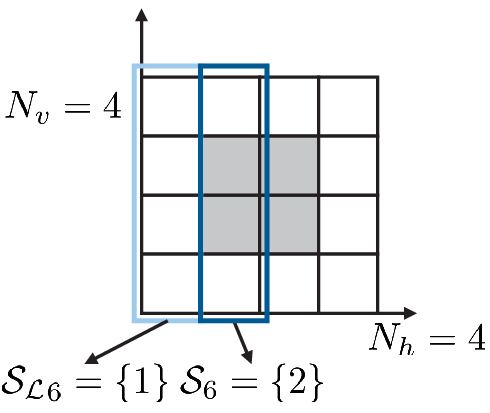}\label{fig6f}}
    \caption{An illustration of our proposed sortPM-based method for estimating $n_{h,{\rm min}}$, where we get a false answer $Y_k$ in the first round of Q\&A.}\label{fig6}
\end{figure}
\begin{figure}
	\centering
	\includegraphics[scale=0.44]{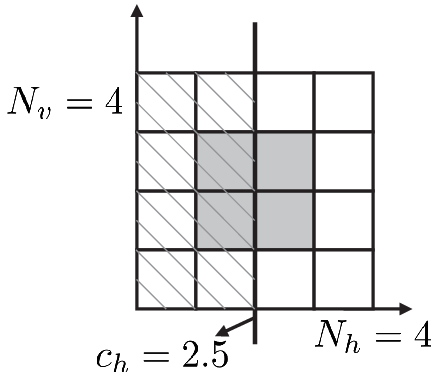}
	\caption{An illustration of our proposed three-phase bisection approach for estimating $n_{h,{\rm min}}$, where we make a false detection at the first time slot.}
	\label{fig7}
\end{figure}

\begin{example}
	In the following, we provide an example to illustrate how the above three-phase bisection approach works to iteratively find $n_{h,{\rm min}}$ and $n_{h,{\rm max}}$ in the horizontal domain. In this example, we assume that on an IRS consisting of $4\times 4$ elements, all the defective IRS elements are located in the region $\mathcal{E}=\{(n_h,n_v)|n_h\in\{2,3\}, n_v\in \{2,3\}\}$, as shown in Fig. \ref{fig5}. Specifically, at the initial stage, we have $n_{h,{\rm min}}^{{\rm lb},0}=n_{h,{\rm max}}^{{\rm lb},0}=1$ and $n_{h,{\rm min}}^{{\rm ub},0}=n_{h,{\rm max}}^{{\rm ub},0}=N_h=4$. As shown in Fig. \ref{fig5a}, at the first time slot, which is Phase I, the boundary to cut in the horizontal domain is $c_h=2.5$, which is useful for updating the bounds for both $n_{h,{\rm min}}$ and $n_{h,{\rm max}}$. Based on the detectors shown in (\ref{eqn:threshold 1})-(\ref{eqn:threshold 3_2}), Case 3 should be detected. In this case, we update $n_{h,{\rm min}}^{{\rm ub},1}=2$, $n_{h,{\rm max}}^{{\rm lb},1}=3$, and keep $n_{h,{\rm min}}^{{\rm lb},1}=n_{h,{\rm min}}^{{\rm lb},0}=1$ and $n_{h,{\rm max}}^{{\rm ub},1}=n_{h,{\rm max}}^{{\rm ub},0}=4$. Because Case 3 has occurred, Phase I ends after the first iteration. As shown in Fig. \ref{fig5b}, at the second time slot, which is Phase II, the boundary to cut in the horizontal domain is $c_h=1.5$, which is useful for checking whether $n_{h,{\rm min}}>1.5$ or $n_{h,{\rm min}}<1.5$. Based on the detectors shown in (\ref{eqn:threshold 1})-(\ref{eqn:threshold 3_2}), Case 2 should be detected. In this case, we update $n_{h,{\rm min}}^{{\rm lb},2}=2$, and keep $n_{h,{\rm min}}^{{\rm ub},2}=n_{h,{\rm min}}^{{\rm ub},1}=2$. Because $n_{h,{\rm min}}^{{\rm lb},2}=n_{h,{\rm min}}^{{\rm ub},2}=2$, Phase II ends and $n_{h,{\rm min}}=2$ can be determined. As shown in Fig. \ref{fig5c}, at the third time slot, which is Phase III, the boundary to cut in the horizontal domain is $c_h=3.5$, which is useful for updating the bounds for $n_{h,{\rm max}}$. Similarly, based on the detectors shown in (\ref{eqn:threshold 1})-(\ref{eqn:threshold 3_2}), Case 1 should be detected. In this case, we update $n_{h,{\rm max}}^{{\rm ub},3}=3$, and keep $n_{h,{\rm max}}^{{\rm lb},3}=n_{h,{\rm max}}^{{\rm lb},1}=3$. Because $n_{h,{\rm max}}^{{\rm lb},3}=n_{h,{\rm max}}^{{\rm ub},3}=3$, Phase III ends and $n_{h,{\rm max}}=3$ can be determined. After three iterations, the values of $n_{h,{\rm min}}$ and $n_{h,{\rm max}}$ are known. We can also apply the above approach to estimate $n_{v,{\rm min}}$ and $n_{v,{\rm max}}$ in the vertical domain.\par
\end{example}
\begin{remark}
	In \textit{Example 2}, at each time slot $t$, our decision about $n_{h,{\rm min}}$ and $n_{h,{\rm max}}$ is always correct. However, because of the estimation errors in $\bar{\boldsymbol{g}}_{\rm e}$ and $\bar{\boldsymbol{g}}_{\rm w}$ as well as the noise at each time slot $t$, sometimes we mistakenly determine which of Case 1, Case 2, and Case 3 is true, and the sub-region containing $n_{h,{\rm min}}$ may be cut off. In the following we assume this false detection is correct, and this sub-region will never be considered. Therefore, the previous false detection cannot be revised in the bisection-based method.\par 
	However, in \textit{Example 3}, we will show that although the answer for $n_{h,{\rm min}}$ is false in some round $k$, the sortPM-based method can still correctly estimate $n_{h,{\rm min}}$. Note that in each round $k$, based on ${Y}_{k-1}$, the posterior probabilities for all columns are updated even if they are really close to 0. Therefore, the previous false answer can be revised by the following right answers in the sortPM-based method.\par
\end{remark}
\begin{example}
In the following, we provide an example to illustrate how the sortPM-based approach works to find $n_{h,{\rm min}}$ in the horizontal domain. In this example, we assume that on an IRS consisting of $4\times 4$ elements, all the defective IRS elements are located in the region $\mathcal{E}=\{(n_h,n_v)|n_h\in\{2,3\}, n_v\in \{2,3\}\}$, as shown in Fig. \ref{fig6}. Under the sortPM-based method, initially, we have $\boldsymbol{\pi}(0)=[0.25,0.25,0.25,0.25]$.\par
As shown in Fig. \ref{fig6a}, in the first round, the query set is defined as $\{1,2\}$ based on $\boldsymbol{\pi}(0)$ according to (\ref{eqn:query set sortPM}) and (\ref{eqn:l star sortPM}). Based on the detectors shown in (\ref{eqn:Case A true}) and (\ref{eqn:Case B true}) for $\mathcal{U}_1$, we get a false answer that $Y_1=0$ because of the estimation errors and the noise in $p(\boldsymbol{y}_{1_1}|{\rm Case~A})$. Then we update $\boldsymbol{\pi}(1)=[0.05,0.05,0.45,0.45]$ based on ${Y}_1$ according to (\ref{eqn:pi k-1 0}).\par
As shown in Fig. \ref{fig6b}, in the second round, the query set $\mathcal{S}_2$ is defined as $\{3\}$ based on $\boldsymbol{\pi}(1)$ according to (\ref{eqn:query set sortPM}) and (\ref{eqn:l star sortPM}). Based on the detectors shown in (\ref{eqn:Case A true}) and (\ref{eqn:Case B true}) for $\mathcal{U}_2$ and $\mathcal{U_L}_2$, we claim that $Y_2=0$. Then, we update $\boldsymbol{\pi}(2)=[0.0833,0.0833,0.0833,0.75]$ based on ${Y}_2$ according to (\ref{eqn:pi k-1 0}).\par
As shown in Fig. \ref{fig6c}, in the third round, the query set $\mathcal{S}_3$ is defined as $\{4\}$ based on $\boldsymbol{\pi}(2)$ according to (\ref{eqn:query set sortPM}) and (\ref{eqn:l star sortPM}). Based on the detectors shown in (\ref{eqn:Case A true}) and (\ref{eqn:Case B true}) for $\mathcal{U}_3$ and $\mathcal{U_L}_3$, we claim that $Y_3=0$. Then, we update $\boldsymbol{\pi}(3)=[0.25,0.25,0.25,0.25]$ based on ${Y}_3$ according to (\ref{eqn:pi k-1 0}).\par
As shown in Fig. \ref{fig6d}, in the forth round, the query set $\mathcal{S}_4$ is defined as $\{1,2\}$ based on $\boldsymbol{\pi}(3)$ according to (\ref{eqn:query set sortPM}) and (\ref{eqn:l star sortPM}). Based on the detectors shown in (\ref{eqn:Case A true}) and (\ref{eqn:Case B true}) for $\mathcal{U}_4$, we claim that $Y_4=1$. Then, we update $\boldsymbol{\pi}(4)=[0.45,0.45,0.05,0.05]$ based on ${Y}_4$ according to (\ref{eqn:pi k-1 1}).\par
As shown in Fig. \ref{fig6e}, in the fifth round, the query set $\mathcal{S}_5$ is defined as $\{1\}$ based on $\boldsymbol{\pi}(4)$ according to (\ref{eqn:query set sortPM}) and (\ref{eqn:l star sortPM}). Based on the detectors shown in (\ref{eqn:Case A true}) and (\ref{eqn:Case B true}) for $\mathcal{U}_5$, we claim that $Y_5=0$. Then, we update $\boldsymbol{\pi}(5)=[0.0833,0.75,0.0833,0.0833]$ based on ${Y}_5$ according to (\ref{eqn:pi k-1 0}).\par
As shown in Fig. \ref{fig6f}, in the sixth round, the query set $\mathcal{S}_6$ is defined as $\{2\}$ based on $\boldsymbol{\pi}(5)$ according to (\ref{eqn:query set sortPM}) and (\ref{eqn:l star sortPM}). Based on the detectors shown in (\ref{eqn:Case A true}) and (\ref{eqn:Case B true}) for $\mathcal{U}_6$ and $\mathcal{U_L}_6$, we claim that $Y_6=1$. Then, we update $\boldsymbol{\pi}(6)=[0.0119,0.9643,0.0119,0.0119]$ based on ${Y}_6$ according to (\ref{eqn:pi k-1 1}). Because $\pi_{2}(6)>1-\epsilon$, the sortPM-based method converges and we correctly guess $n_{h,{\rm min}}=2$ after six rounds of Q\&A. It is observed that $\pi_2(k)$ first decreases because of the false answer $Y_1$, but then increases gradually according to the following right answers. \par
As for the bisection-based approach, as shown in Fig. \ref{fig7}, at the first time slot, because of the estimation errors in $\bar{\boldsymbol{g}}_{\rm e}$ and $\bar{\boldsymbol{g}}_{\rm w}$ as well as the noise in $\boldsymbol{y}_t$, we mistakenly decide that Case 2 is true and cut off the sub-region on the left hand side of the boundary $c_h=2.5$. Specifically, we assume this false detection is correct such that $n_{h,{\rm min}}$ cannot be correctly estimated in the following.
\end{example}
\section{Numerical Results}
In this section, we provide numerical examples to verify the effectiveness of the proposed methods for localizing the defective IRS elements based on over-the-air measurements. We assume that there are $N=1024$ IRS elements, with $N_h=N_v=32$, within which there is a $4\times4$ defective region. Moreover, we assume that the transmitter, receiver, and IRS are deployed in the near-field regime of each other. Therefore, based on the near-field LOS channel model \cite{David2005}, the channel vector from the transmitter to the receiver, i.e., $\boldsymbol{h}$, the channel vector from IRS element $(n_h,n_v)$ to the receiver, i.e., $\boldsymbol{r}_{n_h,n_v}$, and the channel coefficient from the transmitter to IRS element, i.e., $u_{n_h,n_v}$, are given as
\begin{align}
	& \boldsymbol{h}=\alpha^{{\rm TR}}[e^{-j2\pi \|\boldsymbol{s}^{{\rm T}}-\boldsymbol{s}^{{\rm R}}_1\|/\lambda},\nonumber \\ & ~~~~~~~~~~~~ \ldots, e^{-j2\pi \|\boldsymbol{s}^{{\rm T}}-\boldsymbol{s}^{{\rm R}}_M\|/\lambda}]^{T}, \label{eqn:channel 1} \\
	& \boldsymbol{r}_{n_h,n_v}=\alpha^{{\rm IR}}[e^{-j2\pi \|\boldsymbol{s}^{{\rm I}}_{n_h,n_v}-\boldsymbol{s}^{{\rm R}}_1\|/\lambda},\nonumber \\ & ~~~~~~~~~ \ldots, e^{-j2\pi \|\boldsymbol{s}^{{\rm I}}_{n_h,n_v}-\boldsymbol{s}^{{\rm R}}_M\|/\lambda}]^{T}, ~ \forall n_h,n_v,\label{eqn:channel 2} \\
	& u_{n_h,n_v}=\alpha^{{\rm TI}}e^{-j2\pi \|\boldsymbol{s}^{{\rm T}}-\boldsymbol{s}^{{\rm I}}_{n_h,n_v}\|/\lambda}, ~ \forall n_h,n_v,  \label{eqn:channel 3}
\end{align}where $\alpha^{{\rm TR}}$, $\alpha^{{\rm IR}}$, and $\alpha^{{\rm TI}}$ denote the path losses between the transmitter and the receiver, between the IRS and the receiver, and between the transmitter and the IRS, $\boldsymbol{s}^{{\rm T}}$, $\boldsymbol{s}^{{\rm R}}_m$, and $\boldsymbol{s}^{{\rm I}}_{n_h,n_v}$ denote the positions of the transmitter, the $m$-th antenna of the receiver, and IRS element $(n_h,n_v)$, and $\lambda$ denotes the wavelength.\par
We generate the locations of the transmitter, receiver, IRS, and defective elements on the IRS in independent realizations. We decide that the diagnosis is accurate in one realization only when $n_{h,{\rm min}}$, $n_{h,{\rm max}}$, $n_{v,{\rm min}}$, and $n_{v,{\rm max}}$ are all correctly estimated. \par
\begin{figure}
	\centering
	\includegraphics[scale=0.55]{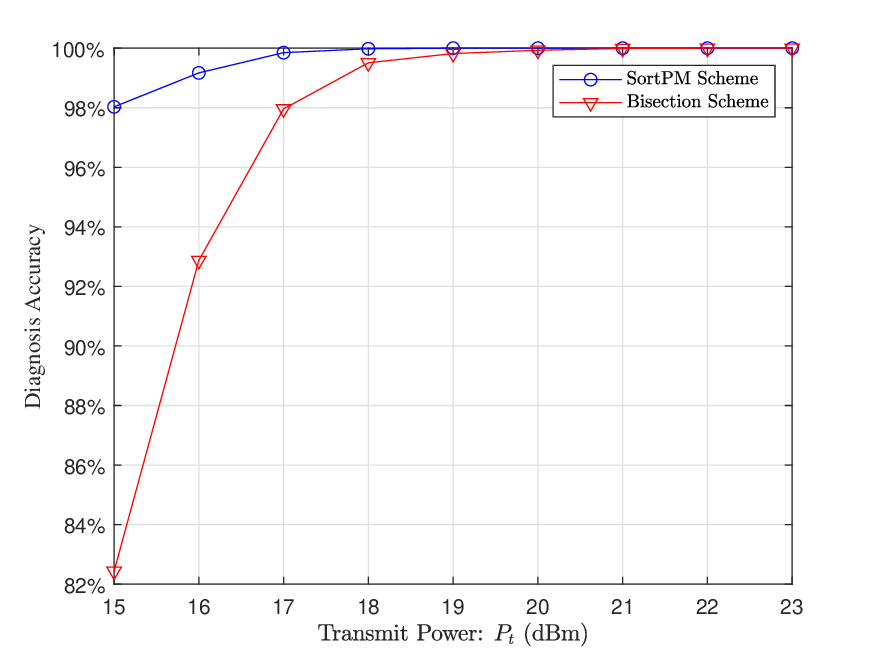}
	\caption{Diagnosis accuracy versus the transmit power $P_{t}$.}
	\label{fig8}
\end{figure}

Fig. \ref{fig8} shows the diagnosis accuracy achieved by the sortPM algorithm and the bisection method versus the transmit power, where the number of receive antennas is $M=4$, and the reliability threshold in Algorithms 1 and 2 is $\epsilon=0.1$. It is observed that in the low SNR regime, the performance of the sortPM method is better than that of the bisection method, while as the SNR increases, the performance gap between these two methods becomes negligible. This is because when the SNR is low, the probability of obtaining wrong answers about whether defective elements exist in some regions of interest is higher. In this case, the sortPM method has a stronger capacity to work well with noisy answers. When the SNR increases, the answer obtained at each Q\&A round tends to be correct with a high probability, and the bisection method can work well in this regime.\par
\begin{figure}
	\centering
	\includegraphics[scale=0.55]{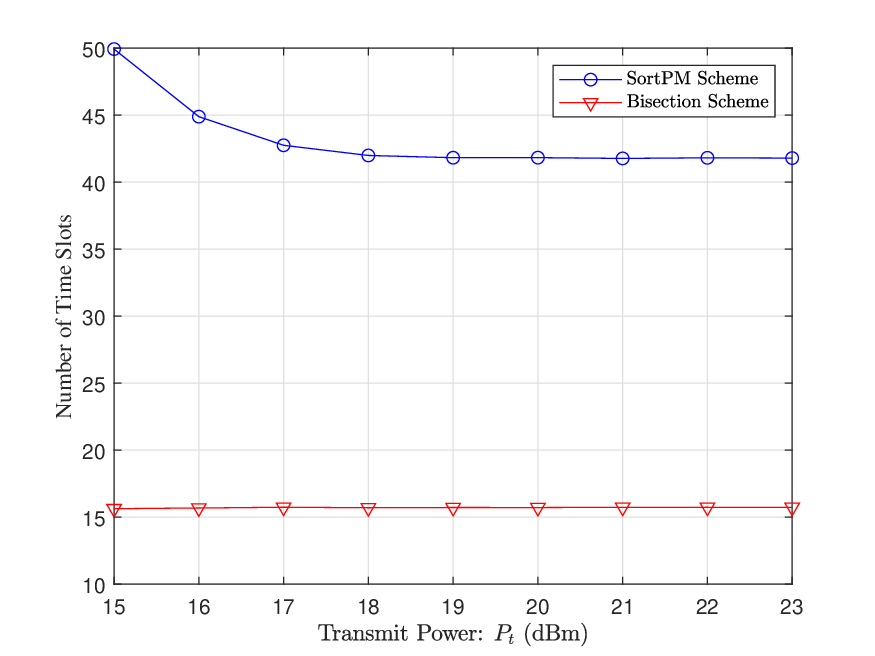}
	\caption{Number of time slots required to find all defective elements versus the transmit power $P_t$.}
	\label{fig9}
\end{figure}

Fig. \ref{fig9} shows the numbers of time slots used by the sortPM and the bisection methods for finding all the defective elements versus the transmit power, where the number of receive antennas is $M=4$, and the reliability threshold in Algorithms 1 and 2 is $\epsilon=0.1$. It is observed that the number of time slots required by the sortPM algorithm decreases with the SNR, while that required by the bisection method is a constant when SNR increases. This is because when SNR increases, the answer about the existence of defective elements in some regions of interest at each Q\&A round tends to be more accurate. In this case, the sortPM method needs a smaller number of Q\&A rounds to correct the previously wrong answers. However, under the bisection method, we remove half of the region to localize the boundary $n_{h,{\rm min}}$, $n_{v,{\rm min}}$, $n_{h,{\rm max}}$, and $n_{v,{\rm max}}$ at each time slot. Therefore, the number of Q\&A rounds has nothing to do with the accuracy of the answer obtained at each Q\&A round. It is also observed that the number of time slots required by the bisection method is significantly smaller than that required by the sortPM algorithm. This is because the sortPM algorithm slowly updates the probabilities of all points to be the boundary point, in case some answers obtained in the previous Q\&A rounds are wrong. \par
\begin{figure}
	\centering
	\includegraphics[scale=0.55]{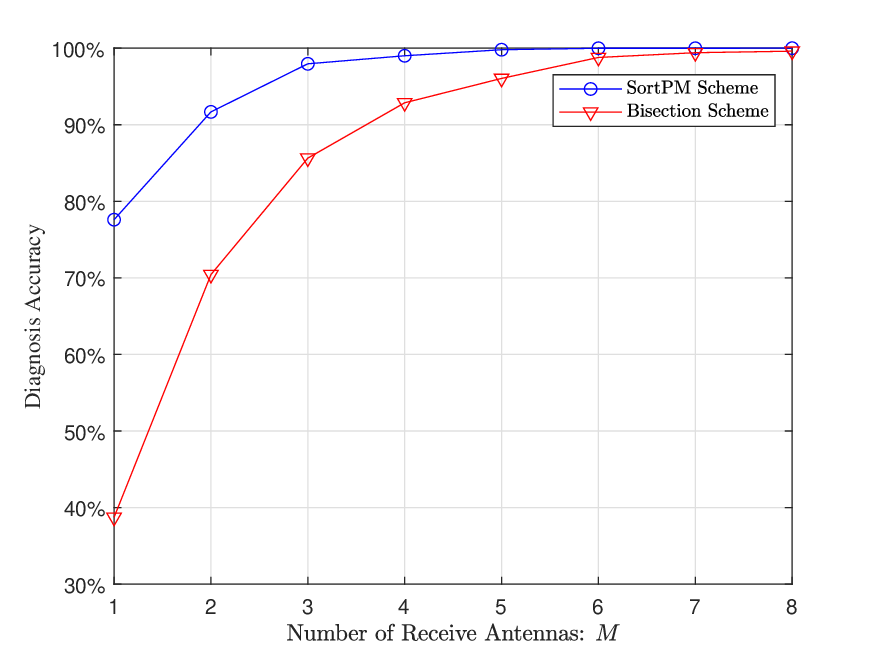}
	\caption{Diagnosis accuracy versus the number of receive antennas $M$.}
	\label{fig10}
\end{figure}
\begin{figure}
	\centering
	\includegraphics[scale=0.55]{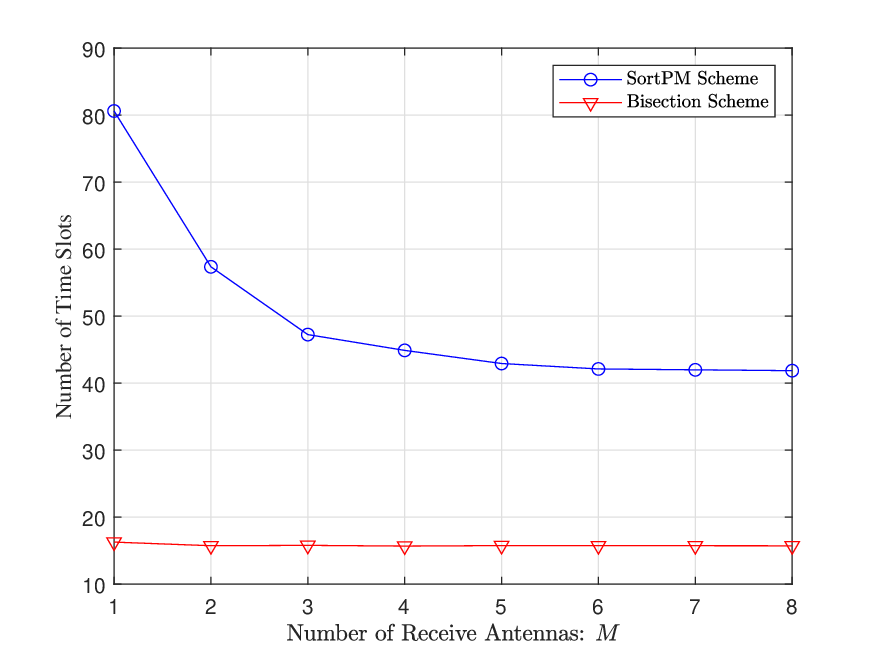}
	\caption{Number of time slots required to find all the defective elements versus the number of receive antennas $M$.}
	\label{fig11}
\end{figure}

Fig. \ref{fig10} shows the diagnosis accuracy achieved by the sortPM method and the bisection method versus the number of receive antennas, where the transmit power is $16~{\rm dBm}$, and the reliability threshold in Algorithms 1 and 2 is $\epsilon=0.1$. It is observed that the diagnosis accuracy increases with the number of receive antennas under both schemes. This is because with more receive antennas, we have more observations to obtain correct answers about whether defective elements exist in some sets of interest at each Q\&A round. It is also observed that the increase in the number of receive antennas can improve diagnosis performance of the bisection method more significantly. This is because the bisection method based diagnosis scheme is more sensitive to wrong answers, and the increase in the number of receive antennas can improve the answer accuracy probability.\par
Fig. \ref{fig11} shows the numbers of time slots requested by the sortPM method and the bisection method versus the number of receive antennas, where the transmit power is $16~{\rm dBm}$, and the reliability threshold in Algorithms 1 and 2 is $\epsilon=0.1$. It is observed that the number of time slots required by the sortPM algorithm decreases with the number of receive antennas, i.e., $M$, while that required by the bisection method is a constant when $M$ increases. This is because when $M$ increases, we have more observations and it is more likely to obtain the correct answer at each Q\&A round. Therefore, under the sortPM method, the number of Q\&A rounds for revising the previously wrong answers decreases with $M$. However, under the bisection method, we assume that the answers obtained in all the Q\&A rounds are always correct. Therefore, the number of Q\&A rounds remains a constant although the answer accuracy probability increases with $M$.
\section{Conclusion}
In this paper, we proposed a novel method for diagnosis of the defective elements on the IRS. This method is simply based on the over-the-air measurements at the receiver, instead of a complicated circuit check process. Under this approach, we showed that the localization of all the defective IRS elements is a 20 questions problem. Two algorithms, i.e., the sortPM algorithm based on noisy 20 questions technique and the bisection algorithm based on  noiseless 20 questions technique, were proposed. To our best knowledge, this work is the first work in the literature to design the over-the-air method for IRS diagnosis.
\bibliographystyle{IEEEtran}
\bibliography{reference}

\begin{thebibliography}{10}
\providecommand{\url}[1]{#1}
\csname url@samestyle\endcsname
\providecommand{\newblock}{\relax}
\providecommand{\bibinfo}[2]{#2}
\providecommand{\BIBentrySTDinterwordspacing}{\spaceskip=0pt\relax}
\providecommand{\BIBentryALTinterwordstretchfactor}{4}
\providecommand{\BIBentryALTinterwordspacing}{\spaceskip=\fontdimen2\font plus
\BIBentryALTinterwordstretchfactor\fontdimen3\font minus
  \fontdimen4\font\relax}
\providecommand{\BIBforeignlanguage}[2]{{%
\expandafter\ifx\csname l@#1\endcsname\relax
\typeout{** WARNING: IEEEtran.bst: No hyphenation pattern has been}%
\typeout{** loaded for the language `#1'. Using the pattern for}%
\typeout{** the default language instead.}%
\else
\language=\csname l@#1\endcsname
\fi
#2}}
\providecommand{\BIBdecl}{\relax}
\BIBdecl

\bibitem{Zhao24}
Z.~Zhao, Z.~Wang, S.~Zhang, and L.~Liu, ``{Finding defective elements in
  intelligent reflecting surface via over-the-air measurements},'' in
  \emph{Proc. IEEE Global Commun. Conf. (GLOBECOM)}, Dec. 2024, pp. 656--661.

\bibitem{Liaskos18}
C.~Liaskos, S.~Nie, A.~Tsioliaridou, A.~Pitsillides, S.~Ioannidis, and
  I.~Akyildiz, ``{A new wireless communication paradigm through
  software-controlled metasurfaces},'' \emph{IEEE Commun. Mag.}, vol.~56,
  no.~9, pp. 162--169, Sep. 2018.

\bibitem{Renzo19}
M.~D. Renzo \emph{et~al.}, ``{Smart radio environments empowered by
  reconfigurable {AI} meta-surfaces: An idea whose time has come},''
  \emph{EURASIP J. Wireless Commun. Netw.}, no. 129, pp. 1--20, May 2019.

\bibitem{Basar19}
E.~Basar, M.~D. Renzo, J.~Rosny, M.~Debbah, M.-S. Alouini, and R.~Zhang,
  ``{Wireless communications through reconfigurable intelligent surfaces},''
  \emph{IEEE Access}, vol.~7, pp. 116\,753--116\,773, 2019.

\bibitem{Qingqing21}
Q.~Wu, S.~Zhang, B.~Zheng, C.~You, and R.~Zhang, ``{Intelligent reflecting
  surface-aided wireless communications: A tutorial},'' \emph{IEEE Trans.
  Commun.}, vol.~69, no.~5, pp. 3313--3351, May 2021.

\bibitem{wang2020channel}
Z.~Wang, L.~Liu, and S.~Cui, ``Channel estimation for intelligent reflecting
  surface assisted multiuser communications: Framework, algorithms, and
  analysis,'' \emph{IEEE Trans. Wireless Commun.}, vol.~19, no.~10, pp.
  6607--6620, Oct. 2020.

\bibitem{liu2020matrix}
H.~Liu, X.~Yuan, and Y.-J.~A. Zhang, ``Matrix-calibration-based cascaded
  channel estimation for reconfigurable intelligent surface assisted multiuser
  {MIMO},'' \emph{IEEE J. Sel. Areas Commun.}, vol.~38, no.~11, pp. 2621--2636,
  Jul. 2020.

\bibitem{Zhou22}
G.~Zhou, C.~Pan, H.~Ren, P.~Popovski, and A.~L. Swindlehurst, ``{Channel
  estimation for RIS-aided multiuser millimeter-wave systems},'' \emph{IEEE
  Trans. Signal Process.}, vol.~70, pp. 1478--1492, 2022.

\bibitem{Chen23}
J.~Chen, Y.-C. Liang, H.~V. Cheng, and W.~Yu, ``{Channel estimation for
  reconfigurable intelligent surface aided multi-user MIMO systems},''
  \emph{IEEE Trans. Wireless Commun.}, vol.~22, no.~10, pp. 6853--6869, Oct.
  2023.

\bibitem{Runnan24}
R.~Liu, L.~Liu, Y.~Xu, D.~He, W.~Zhang, and C.~W. Chen, ``Detecting abrupt
  change of channel covariance matrix in {IRS}-assisted communication,''
  \emph{IEEE Wireless Commun. Lett.}, vol.~13, no.~2, pp. 318--322, Feb. 2024.

\bibitem{Rui25}
R.~Wang, Z.~Wang, L.~Liu, S.~Zhang, and S.~Jin, ``Reducing channel estimation
  and feedback overhead in {IRS}-aided downlink system: {A}
  quantize-then-estimate approach,'' \emph{IEEE Trans. Wireless Commun.},
  vol.~24, no.~2, pp. 1325--1338, Feb. 2025.

\bibitem{Wu19IRS}
Q.~Wu and R.~Zhang, ``Intelligent reflecting surface enhanced wireless network
  via joint active and passive beamforming,'' \emph{IEEE Trans. Wireless
  Commun.}, vol.~18, no.~11, pp. 5394--5409, Nov. 2019.

\bibitem{Mu20IRS}
X.~Mu, Y.~Liu, L.~Guo, J.~Lin, and N.~Al-Dhahir, ``Exploiting intelligent
  reflecting surfaces in {NOMA} networks: Joint beamforming optimization,''
  \emph{IEEE Trans. Wireless Commun.}, vol.~19, no.~10, pp. 6884--6898, Oct.
  2020.

\bibitem{Liu24ISAC}
L.~Liu, S.~Zhang, and S.~Cui, ``Leveraging a variety of anchors in cellular
  network for ubiquitous sensing,'' \emph{IEEE Commun. Mag.}, vol.~62, no.~9,
  pp. 98--104, Sept. 2024.

\bibitem{Zheng19ISAC}
L.~Zheng, M.~Lops, Y.~C. Eldar, and X.~Wang, ``Radar and communication
  coexistence: An overview: A review of recent methods,'' \emph{IEEE Signal
  Process. Mag.}, vol.~36, no.~5, pp. 85--99, Sept. 2019.

\bibitem{Liuan22ISAC}
A.~Liu \emph{et~al.}, ``A survey on fundamental limits of integrated sensing
  and communication,'' \emph{IEEE Commun. Surveys Tut.}, vol.~24, no.~2, pp.
  994--1034, 2nd Quta. 2022.

\bibitem{Song2022ISAC}
X.~Song, D.~Zhao, H.~Hua, T.~X. Han, X.~Yang, and J.~Xu, ``Joint transmit and
  reflective beamforming for {IRS}-assisted integrated sensing and
  communication,'' in \emph{Proc. IEEE Wireless Commun. Netw. Conf. (WCNC)},
  Apr. 2022, pp. 189--194.

\bibitem{Shao2022ISAC}
X.~Shao, C.~You, W.~Ma, X.~Chen, and R.~Zhang, ``Target sensing with
  intelligent reflecting surface: Architecture and performance,'' \emph{IEEE J.
  Sel. Areas Commun.}, vol.~40, no.~7, pp. 2070--2084, Jul. 2022.

\bibitem{Fang2024ISAC}
Y.~Fang, S.~Zhang, X.~Li, X.~Yu, J.~Xu, and S.~Cui, ``Multi-{IRS}-enabled
  integrated sensing and communications,'' \emph{IEEE Trans. Commun.}, vol.~72,
  no.~9, pp. 5853--5867, Sept. 2024.

\bibitem{Srinivasan04}
J.~Srinivasan, S.~V. Adve, P.~Bose, and J.~A. Rivers, ``The case for lifetime
  reliability-aware microprocessors,'' in \emph{Proc. 31st Annu. Int. Symp.
  Comput. Archit.}, Jun. 2004, pp. 276--287.

\bibitem{Saeed18}
T.~Saeed \emph{et~al.}, ``{Fault adaptive routing in metasurface controller
  networks},'' in \emph{Int. Workshop Netw. Chip Archit. (NoCArc)}, Oct. 2018,
  pp. 1--6.

\bibitem{Taghvaee20}
H.~Taghvaee, A.~Cabellos-Aparicio, J.~Georgiou, and S.~Abadal, ``{Error
  analysis of programmable metasurfaces for beam steering},'' \emph{IEEE J.
  Emerg. Sel. Top. Circuits Syst.}, vol.~10, no.~1, pp. 62--74, Mar. 2020.

\bibitem{Badiu20}
M.~Badiu and J.~P. Coon, ``{Communication through a large reflecting surface
  with phase errors},'' \emph{IEEE Wireless Commun. Lett.}, vol.~9, no.~2, pp.
  184--188, Feb. 2020.

\bibitem{Khel22}
A.~M.~T. Khel and K.~A. Hamdi, ``{Effects of hardware impairments on
  {IRS}-enabled {MISO} wireless communication systems},'' \emph{IEEE Commun.
  Lett.}, vol.~26, no.~2, pp. 259--263, Feb. 2022.

\bibitem{Alouini21}
X.~Qian, M.~D. Renzo, J.~Liu, A.~Kammoun, and M.-S. Alouini, ``{Beamforming
  through reconfigurable intelligent surfaces in single-user MIMO systems: SNR
  distribution and scaling laws in the presence of channel fading and phase
  noise},'' \emph{IEEE Wireless Commun. Lett.}, vol.~10, no.~1, pp. 77--81,
  Jan. 2021.

\bibitem{Zhang22}
J.~Zhang, X.~Hu, and C.~Zhong, ``{Phase calibration for intelligent reflecting
  surfaces assisted millimeter wave communications},'' \emph{IEEE Trans. Signal
  Process.}, vol.~70, pp. 1026--1040, 2022.

\bibitem{Bucci05}
O.~M. Bucci, M.~D. Migliore, G.~Panariello, and P.~Sgambato, ``Accurate
  diagnosis of conformal arrays from near-field data using the matrix method,''
  \emph{IEEE Trans. Antennas Propag.}, vol.~53, no.~3, pp. 1114--1120, Mar.
  2005.

\bibitem{Migliore11}
M.~D. Migliore, ``A compressed sensing approach for array diagnosis from a
  small set of near-field measurements,'' \emph{IEEE Trans. Antennas Propag.},
  vol.~59, no.~6, pp. 2127--2133, Jun. 2011.

\bibitem{Oliveri12}
G.~Oliveri, P.~Rocca, and A.~Massa, ``Reliable diagnosis of large linear
  arrays—a bayesian compressive sensing approach,'' \emph{IEEE Trans.
  Antennas Propag.}, vol.~60, no.~10, pp. 4627--4636, Oct. 2012.

\bibitem{Fuchs16}
B.~Fuchs, L.~L. Coq, and M.~D. Migliore, ``Fast antenna array diagnosis from a
  small number of far-field measurements,'' \emph{IEEE Trans. Antennas
  Propag.}, vol.~64, no.~6, pp. 2227--2235, Jun. 2016.

\bibitem{Xiong19}
C.~Xiong, G.~Xiao, Y.~Hou, and M.~Hameed, ``A compressed sensing-based element
  failure diagnosis method for phased array antenna during beam steering,''
  \emph{IEEE Antennas Wireless Propag. Lett.}, vol.~18, no.~9, pp. 1756--1760,
  Sept. 2019.

\bibitem{Renyi1984}
A.~R{\'e}nyi and Z.~Makkai-Bencs{\'a}th, \emph{A diary on information
  theory}.\hskip 1em plus 0.5em minus 0.4em\relax Akademiai Kiado Budapest,
  1984.

\bibitem{Ulam1991}
S.~M. Ulam, \emph{Adventures of a Mathematician}.\hskip 1em plus 0.5em minus
  0.4em\relax Univ of California Press, 1991.

\bibitem{Schalk1971}
J.~Schalkwijk, ``A class of simple and optimal strategies for block coding on
  the binary symmetric channel with noiseless feedback,'' \emph{IEEE Trans.
  Inf. Theory}, vol.~17, no.~3, pp. 283--287, May 1971.

\bibitem{Hill1992}
R.~Hill and J.~Karim, ``Searching with lies: the {Ulam} problem,''
  \emph{Discrete Math.}, vol. 106-107, pp. 273--283, 1992.

\bibitem{Pelc02}
A.~Pelc, ``Searching games with errors—fifty years of coping with liars,''
  \emph{Theor. Comput. Sci.}, vol. 270, no.~1, pp. 71--109, 2002.

\bibitem{Chiu16}
S.-E. Chiu and T.~Javidi, ``{Sequential measurement-dependent noisy search},''
  in \emph{Proc. IEEE Inf. Theory Workshop}, Sept. 2016, pp. 221--225.

\bibitem{Chiu21}
------, ``{Low complexity sequential search with size-dependent measurement
  noise},'' \emph{IEEE Trans. Inf. Theory}, vol.~67, no.~9, pp. 5731--5748,
  Sept. 2021.

\bibitem{Chiu19}
S.-E. Chiu, \emph{Noisy binary search: Practical algorithms and
  applications}.\hskip 1em plus 0.5em minus 0.4em\relax University of
  California, San Diego, 2019.

\bibitem{David2005}
D.~Tse and P.~Viswanath, \emph{Fundamentals of Wireless Communication}.\hskip
  1em plus 0.5em minus 0.4em\relax Cambridge, UK: Cambridge University Press,
  2005.

\end{thebibliography}

\end{document}